\documentstyle[prl,preprint,aps,epsfig]{revtex}
\tightenlines

\begin{document}

\def\mev{\hbox{\ MeV}}
\def\kev{\hbox{\ keV}}

\def\TG{{{\tilde\Gamma}}}
\def\TE{{\tilde E}}
\def\TW{\tilde W}
\def\CH{{\cal H}}
\def\TCE{\tilde {\cal E}}
\def\Tg{\tilde \gamma}
\def\TP{\tilde \Phi}

\draft

\title{Effective Hamiltonian and unitarity of the $S$ matrix}

\author{
I. Rotter }
\address{Max-Planck-Institut f\H{u}r Physik komplexer Systeme, D-01187
Dresden, Germany 
}

\date{\today}

\maketitle

\begin{abstract}

The properties of  open quantum systems are described well by an effective 
Hamiltonian ${\cal H}$ 
that consists of two parts: the Hamiltonian  $H$ of the closed system with 
discrete eigenstates and the coupling matrix
$W$ between discrete states and continuum.
The eigenvalues of ${\cal H}$ determine the poles of the $S$ matrix.  
The coupling matrix elements $\tilde W_k^{cc'}$ between the  
eigenstates $k$ of ${\cal H}$ and the continuum may be very different from the 
coupling matrix elements $W_k^{cc'}$ 
between the eigenstates of $H$ and the continuum.
Due to the unitarity of the $S$ matrix, the $\TW_k^{cc'}$
depend on energy in a non-trivial manner, 
that  conflicts with the assumptions of  some  approaches
to reactions in the overlapping regime. 
Explicit expressions for  the wave functions of the 
resonance states and for their phases in the neighbourhood of, 
respectively, avoided level crossings in the complex plane 
and double poles of the $S$ matrix  are given.

\end{abstract}

\bigskip
\pacs{PACS numbers: 73.21.La,  72.15.Qm,  03.65.Nk,  03.65.Vf
}

\newpage

\section{Introduction}
 
Quantum systems are characterized by a number of discrete states the structure
of which is more or less complicated. These quantum systems do, however, 
not exist isolated from other systems. Most of them
are  embedded in an environment, e.g. in the continuum of decay channels.
System and environment interact with one another, and 
it is this interaction that 
allows us to study the properties
of the system. The feedback of the interaction with the environment 
onto the properties of the
system itself is an old problem raised in the very beginning of     
quantum mechanics. It becomes the more important the smaller the system
is. 

As an example, most states of a nucleus are embedded in the contiuum of decay
channels due to which they get a finite life time. In other words: the
discrete states of a nucleus shade off into resonance states with complex
energies ${\cal E}_k = E_k - \frac{i}{2} \, \Gamma_k$. The  $E_k$ give the 
positions in energy of the resonance states while the 
widths  $\Gamma_k$ are characteristic of their life times. The $E_k$ may be
different from the energies of the discrete states, and the widths 
$\Gamma_k$ may be large corresponding to a short life time. Nevertheless,
there is a well defined relation between the 
discrete states characterizing  the closed system, and the resonance states
appearing in the open system. 
The main difference in the theoretical description 
of quantum systems without and with coupling to an environment is that the
function space of the 
system is supposed to be complete in the first case
while this is not so in the second case. 
Accordingly, the Hamilton operator is Hermitian in the first case, and
the eigenvalues are discrete.
The resonance states, however, characterize a subsystem
described by a non-Hermitian Hamilton operator with complex eigenvalues. The
function space containing everything consists, in the second case, of system 
plus environment.

The mathematical formulation of this problem goes back to Feshbach 
\cite{feshbach} who
introduced the two subspaces $Q$ and $P$, with $Q+P=1$, containing the discrete
and scattering states, respectively. Feshbach was able to formulate a
unified description of nuclear reactions with  direct
processes in the short-time scale and  compound nucleus processes
in the long-time scale. Due to the high 
excitation energy and high level
density in compound nuclei, he introduced statistical approximations 
in order to describe the
discrete states of the $Q$ subspace. A unified description of nuclear
structure and nuclear reaction aspects is much more complicated and became
possible only at the end of the last century (see \cite{rep} for a recent
review). In this formulation, the states of both subspaces are described with
the same accuracy. All the coupling matrix elements between different discrete
states, different scattering states as well as between discrete and scattering 
states have to be calculated in order to get results that can be compared with
experimental data.   This method has been applied to the description of light
nuclei by using the shell model approach for  the
discrete many-particle states of the $Q$ subspace \cite{rep}. 

In the unified description of structure and reaction aspects, 
the system is described by an effective  Hamiltonian ${\cal H}$ that 
consists of two terms: the Hamiltonian matrix $H$ of the closed system 
with discrete eigenstates 
and the coupling matrix  between system and environment. 
The last term is responsible for 
the finite lifetime of the resonance states. The
eigenvalues of ${\cal H}$ are complex and give the poles of the $S$ matrix.
The motion of these eigenvalues  as a function of a certain parameter
is discussed in many papers (see the recent review \cite{rep}).
The statistics of complex eigenvalues and the corresponding nonorthogonal
eigenvectors for non-Hermitian random matrices are recently considered in
\cite{mehlig}.

The coupling matrix  elements $\TW_k^{cc'}$
between the resonance states and the 
continuum are seldom studied. Their relation to the
coupling matrix elements  $W_k^{cc'}$ between 
the corresponding discrete states and the continuum
can be expressed by the mixing coefficients that appear in the
representation of the wave functions of the resonance states in the
set of wave functions of the discrete states. 
Generically, the relation between the wave functions of the resonance 
states and 
those of the discrete states is complicated since the number of discrete
states of a realistic system is large. Many of them can contribute
to the wave function of a certain resonance state, 
almost independently of their energetical distance, see e.g. \cite{jung}. 
The numerical results for the coupling matrix  elements $\TW_k^{cc'}$
of nuclear states show a non-trivial energy 
dependence, especially at high level density  \cite{rep}. 
In the statistical approach  to nuclear  reactions 
and  the  application of this approach to some other reactions they are, 
however, assumed to be simple, energy-independent parameters
such as the  $W_k^{cc'}$,  e.g. \cite{guhr,schanz,fedorov,dittes,savin}.

It is the aim of the present paper (i) to study the energy dependence of 
the coupling matrix elements $\TW_k^{cc'}$ between resonance states 
and continuum 
that follows immediately from the unitarity of the $S$ matrix, and  (ii) 
to study the behaviour of the wave 
functions of the resonance states in the overlapping regime
since they determine the energy dependence of the $\TW_k^{cc'}$
in numerical calculations performed for special systems. Most 
interesting is the behaviour of the 
wave functions near avoided level crossings in the complex plane.

As the results show, the coupling coefficients  $\TW_k^{cc'}$
are, generically, energy dependent. The energy
dependence is, however, not important as long as the distance
in energy  between the resonance states is larger than the sum of their 
widths. This result is in full agreement with the 
statement of the authors of the review \cite{guhr} who restricted the
application of their approach  to the non-overlapping regime. 
In the overlapping regime, however, the energy dependence of the coupling 
coefficients $\TW_k^{cc'}$ can not be neglected. It is a consequence of  
the unitarity of the $S$ matrix and causes 
nonlinear terms in the $S$ matrix at high level density. 
Nevertheless, the line shape of the resonances can equivalently be
described by the energy independent $W_k^{cc'}$ in many cases. 
The $W_k^{cc'}$ lose, however, their physical meaning in the overlapping 
regime \cite{marost4a}. 
As to the wave functions  of two resonance states $k$ and $l$
in the neighbourhood 
of an avoided level crossing in the complex plane, they are mixed:
$\beta_k \, \TP_k \pm i \,\beta_l \, \TP_l$.
The corresponding phase changes in approaching the critical value
of the parameter at which the levels avoid crossing (where
$\beta_k = \pm \beta_l$), 
are caused by nonlinear terms in the Schr\"odinger equation.  
These terms are, finally, responsible for the energy
dependence of the coupling coefficients $\TW_k^{cc'}$  
between system and continuum in numerical calculations. This result
following from the behaviour of the wave functions of the resonance
states at high level density,
coincides with that following from the unitarity of the $S$ matrix. 

The paper is organized in the following manner. In Sect. II, the 
main ingredients of the unified description of structure and reaction
aspects of a quantum system embedded in a continuum are given. 
These are the effective Hamiltonian ${\cal H}$ of the system 
and the $S$ matrix which both are derived by solving the 
Schr\"odinger equation in the full function 
space with discrete and continuous states. Further, some properties of the 
spectroscopic values that characterize the system are sketched
with special emphasis of their behaviour in the overlapping regime. 
In Sect. III, 
the coupling coefficients $\TW$ between system and continuum 
are directly obtained in the one-channel case by starting
from a unitary  $S$ matrix. The nonlinear effects  appearing 
in the overlapping regime are discussed. The wave functions  
at avoided level crossings in the complex plane and in its neighbourhood 
are derived in Sect. IV. They are energy dependent 
and their phases change in a certain range around the critical 
value of the parameter at which the resonance states avoid crossing. 
Sect. V contains concluding remarks on the energy dependence 
of the coupling coefficients $\TW_k^{cc'}$ appearing in a model
with unified  description of structure and reaction aspects
at high level density.

\section{Effective Hamiltonian and 
$S$ matrix for a quantum system embedded in a continuum}

In the unified description of structure and reaction aspects of quantum 
systems, the Schr\"odinger equation 
\begin{eqnarray} 
(H^{\rm full} -E)\Psi (E) = 0
\label{csm1}
\end{eqnarray}
is solved in
a function space containing everything, i.e. discrete as well as continuous
states. The Hamilton operator $H^{\rm full}$ is Hermitian, 
the wave functions $\Psi$ 
depend on energy as well as on the decay channels and all the resonance
states of the system. Knowing the wave functions 
$\Psi (E)$, an expression for the $S$ matrix can be derived
that holds true also in the overlapping regime,
see the recent review \cite{rep}. It reads
\begin{eqnarray} 
 S_{cc'} = e^{i(\delta_c-\delta_{c'})}\; \Big[
 \delta_{cc'} -  S_{cc'}^{(1)} -  S_{cc'}^{(2)} \Big] ~ \ ,
\label{csm2}
\end{eqnarray}
where   $\delta_c$ is the phase shift in channel $c$, $~S_{cc'}^{(1)}$
is the smooth direct reaction part related to the short-time scale,
and 
\begin{eqnarray}
 S_{cc'}^{(2)} =  i \; \sum_{k=1}^N
  \frac{\tilde\gamma_{k}^{c}\; \tilde\gamma_{k}^{c'}}
  {E - {\tilde E}_k + \frac{i}{2} {\tilde \Gamma}_k} 
\label{csm3}
\end{eqnarray}
is the resonance reaction part related to the long-time scale.
Here, the   $\tilde{\cal E}_k = \tilde E_k - \frac{i}{2}\; 
\tilde\Gamma_k$ are the complex eigenvalues 
of the non-Hermitian Hamilton operator 
\begin{eqnarray}
{\cal H} \equiv {\cal H}_{QQ} = H_{QQ} + H_{QP} \, G_P^{(+)} \, H_{PQ} 
\label{csm4}
\end{eqnarray}
appearing effectively in the system ($Q$ subspace)
after embedding it into the continuum ($P$ subspace). It is $H^{\rm full}
\equiv H_{QQ} + H_{QP} + H_{PQ} + H_{PP}$  where $H_{QQ}$ is the
Hamiltonian of the closed system and $H_{PP}$ is that 
for the environment (scattering states). The two 
terms $H_{PQ}$ and $H_{QP}$ characterize the coupling between the two 
subspaces. They appear in the source terms of the equations 
in either subspace when  $Q+P=1$ as well as in the
effective Hamiltonians of the subspaces.  The
effective Hamiltonian in the $Q$ subspace is given in (\ref{csm4}),
and an analogous expression for the effective Hamiltonian in the 
$P$ subspace can be written down, see \cite{rep}.
Usually, Re$\,(H_{QP} \, G_P^{(+)} \, H_{PQ}) \ne 0$ and Re$\,({\cal H}) 
\equiv  H_{QQ} + {\rm Re}\,(H_{QP} \, G_P^{(+)} \, H_{PQ}) \ne   H_{QQ}$,
see \cite{rep} for a detailed discussion. 
The $G_P^{(+)}$ in (\ref{csm4}) are the Green functions in the $P$ 
subspace, and Im$\,(H_{QP} \, G_P^{(+)} \, H_{PQ})$ is determined by
the coupling matrix elements 
\begin{eqnarray}
\tilde\gamma_k^c = 
\sqrt{2\pi} \; \langle \xi_E^c | V | \tilde \Phi_k \rangle
\label{over1}
\end{eqnarray}
between the resonance states and the scattering states.
The   $\tilde{\cal E}_k$ are energy dependent functions. 
They are complex and also energy dependent functions. In (\ref{over1}),
the $\xi_E^c$ are the scattering wave functions and the 
$\tilde \Phi_k$ are the eigenfunctions of ${\cal H}$, Eq. (\ref{csm4}). 
The wave functions $\tilde\Omega_k$ of the resonance states 
are related to the eigenfunctions $\tilde \Phi_k$ of ${\cal H}$ by a
Lippmann-Schwinger like relation \cite{rep},
\begin{eqnarray}
\tilde \Omega_k = (1+  G_P^{(+)} H_{PQ}) \, \tilde \Phi_k
\; .
\label{csm4a}
\end{eqnarray}
The eigenfunctions of ${\cal H}$
are bi-orthogonal,
\begin{eqnarray} 
\langle \tilde \Phi_l^* | \tilde \Phi_k \rangle = \delta_{kl} 
\label{csm5}
\end{eqnarray}
so that
\begin{eqnarray}
& & \langle \tilde \Phi_k|\tilde\Phi_k \rangle =
{\rm Re} ( \langle \tilde\Phi_k|\tilde\Phi_k \rangle) 
 \; \;  ;  \; \; \;
 A_k \equiv   \langle \tilde \Phi_k|\tilde\Phi_k \rangle
\ge 1             
 \label{csm6}
\end{eqnarray}
and 
\begin{eqnarray}
& & \langle \tilde\Phi_k|\tilde\Phi_{l\ne k} \rangle  =  
i \; {\rm Im} (\langle \tilde\Phi_k|\tilde\Phi_{l\ne k} \rangle )  
 =  -\langle \tilde\Phi_{l\ne k}|\tilde\Phi_k \rangle 
\; \;  ;  \; \; B_k^{l\ne k} \equiv  
| \langle\tilde \Phi_k|\tilde\Phi_{l\ne k} \rangle| \ge 0 ~ \ . 
  \label{csm7}
\end{eqnarray}
It should be noticed that the standard normalization 
$\langle \tilde \Phi_l | \tilde \Phi_k \rangle = \delta_{kl}$ is
equivalently to Eqs. (\ref{csm5}) up to   (\ref{csm7}) for all $k,l$ 
but those with 
$\tilde{\cal E}_k =  \tilde{\cal E}_l$ (double pole of the $S$ matrix)
where $A_k \to \infty$ and $B_k^l \to \infty$.
As a consequence of (\ref{csm6}) holds \cite{rep}
\begin{eqnarray}
\tilde \Gamma_k = \frac{\sum_c |\tilde \gamma_k^c|^2}{ A_k}
\; \le \;  \sum_c |\tilde \gamma_k^c|^2 ~ \ .
\label{csm8}
\end{eqnarray}

The main difference to the standard theory is that
the $\tilde\Gamma_k, \; \tilde \gamma_k^c$ and $\tilde E_k$
are not numbers but  energy dependent functions
\cite{rep}.  The energy dependence of  
${\rm Im}\{\tilde {\cal E}_k\} =  - \frac{1}{2} \, \tilde \Gamma_k$
is large near to the  threshold for opening the first decay channel. 
This causes not only deviations from the Breit Wigner line shape
of isolated resonances lying near to the threshold, but also an 
interference with the above-threshold ''tail'' of bound states  \cite{rep}. 
Also an inelastic threshold may have an influence on the line shape
of a resonance when the resonance  lies near to the threshold and  
is coupled strongly to the channel which opens \cite{ro91}.
Also in this case, $\tilde \Gamma_k$ depends strongly on energy.  
In the cross section, a cusp may appear 
in the cross section instead of a resonance  of Breit Wigner  shape. Both
types of threshold effects in the line shape of resonances  
can explain experimental data known in nuclear physics \cite{rep}.
They can not be simulated by a parameter in the $S$ matrix.

The energy dependence of $\TE_k$ and $\TG_k$ may be important also far from
decay thresholds \cite{rep}.
Characteristic of the motion of the poles of the $S$ matrix as a function of 
a certain parameter (which may be also the energy $E$ of the system)  
are the following generic results
obtained for very different systems in the overlapping regime: 
the trajectories of the $S$ matrix poles avoid crossing 
with the only exception of exact crossing
when the $S$ matrix has a double (or multiple) pole. At the avoided crossing, 
either level repulsion or level attraction occurs. The first case is caused by
a predominantly real interaction between the crossing states and is 
accompanied by the tendency
to form a uniform  time scale of the system. Level attraction occurs,
however, when the interaction is dominated by its imaginary part 
arising from the coupling via the continuum. It is
accompanied by the formation of different time scales in the system: while
some of the states decouple more or less completely from the continuum 
and become
long-lived (trapped), a few of the states become short-lived and wrap
the long-lived ones in the cross section.
The dynamics of  quantum systems at high level density is determined 
by the interplay of
these two opposite tendencies. For a more detailed discussion see \cite{rep}.

One of these tendencies, the phenomenon of resonance trapping 
\begin{eqnarray}
\sum_{k=1}^N \tilde \Gamma_k \approx \sum_{K=1}^K \tilde \Gamma_k \quad ;
\quad \quad \sum_{k=K+1}^N \tilde \Gamma_k \approx 0 
\; ,
\label{corr1}
\end{eqnarray}
appears only in the overlapping regime. It is caused by Im$(\CH)$ 
and means almost complete decoupling of $N-K$ resonance states from the
continuum while $K$ of them become short-lived  \cite{rep}. 
Usually, $K\ll N-K$. The
long-lived resonance states in the overlapping regime appear often to be well
isolated from one another \cite{marost4a}. The few short-lived resonance 
states determine the evolution of the system. That means,
quick direct reaction processes may appear,
at large overall coupling strength, from slow resonance processes
by means of the resonance trapping phenomenon.
Meanwhile, the phenomenon of  resonance trapping has been
proven experimentally on a microwave cavity as a function of the degree of
opening of the cavity to an attached lead  \cite{stm}. 
In this experiment, the parameter varied is the overall
coupling strength between discrete and scattering states. 
Resonance trapping may appear, however,  as  function of any parameter
\cite{rep}.

In any case, the energies and widths 
of the resonance states follow from the solutions of the fixed-point
equations :
\begin{eqnarray}
E_k = \tilde E_k
{\mbox{\footnotesize $(E\!\!=\!\!E_k)$}}  \qquad {\rm and} \qquad
\Gamma_k  =  \tilde \Gamma_k
 {\mbox{\footnotesize $(E\!\!=\!\!E_k)$}} ~ \ ,
\label{eq:fixp2}
\end{eqnarray}
on condition that the two subspaces are defined adequately \cite{rep}.
The values $E_k$ and $ \Gamma_k$ 
correspond to the standard spectroscopic observables. 
The wave functions of the resonance 
states are defined by the functions 
$\tilde \Omega_k$, Eq. (\ref{csm4a}), at the energy $E=E_k$.
The partial widths are related to the coupling matrix elements 
$(\tilde \gamma_{k}^{c})^2$ that are calculated  independently by means 
of the   eigenfunctions $\tilde \Phi_k$ of ${\cal H}$. 
For isolated resonances, $A_k=1$ according to (\ref{csm6}),  and
the standard relation 
$\Gamma_k = \sum_c |\gamma_{k}^{c}|^2 $ follows from (\ref{csm8}).
In the overlapping regime, the partial widths lose their physical meaning,
since $A>1$. 
Both functions,  $(\tilde\gamma_{k}^{c})^2$ and $\tilde \Gamma_k$,
may even show  a different energy dependence \cite{rep}.

It follows immediately from (\ref{csm4}), that the
coupling of the resonance states via the continuum 
induces additional correlations between the states. These correlations  
are described by
the term $H_{QP}G_P^{(+)}H_{PQ} $ of the effective 
Hamiltonian ${\cal H}$. The real part Re$(H_{QP}G_P^{(+)}H_{PQ})$ 
causes level repulsion
in energy and is accompanied by the tendency to form a uniform time scale 
in the system. In contrast to this behaviour, the imaginary part 
Im$(H_{QP}G_P^{(+)}H_{PQ})$ causes different time scales 
in the system and is accompanied by level attraction in energy. Thus, 
an essential part of the energy dependence of the  eigenvalues of $\CH$
is caused by the additional correlations of the states via the continuum.
They are important, above all, at high level density.

For  isolated resonance states, the additional shift in energy
is usually taken into account by simulating  
Re$({\cal H}) = H_{QQ} + {\rm Re}\, (H_{QP}G_P^{(+)}H_{PQ})$  
by $H_0 + V'$,  where $V'$ is assumed to be describable by
two-body effective residual forces. 
It should be mentioned, however, that  Re$(H_{QP}G_P^{(+)}H_{PQ})$ 
can not completely be simulated 
by an additional contribution to the residual 
two-body interaction even in the case of well isolated resonances, 
since it contains many-body effects,
as follows from the analytical structure of $H_{QP}G_P^{(+)}H_{PQ}$.
Re$(H_{QP}G_P^{(+)}H_{PQ})$ is an integral over energy 
and depends explicitly on the energies $\epsilon_c$ at which  the
channels $c$ open. For details see \cite{rep}.

Spectroscopic studies in the overlapping regime are more complicated.
The  wave functions $\TP_k$ may be represented in the set of eigenfunctions 
$\{\Phi_k\}$ of the Hermitian Hamilton operator  $H \equiv H_{QQ}$ , 
\begin{eqnarray}
\tilde \Phi_k = \sum_l \, b_{kl} \; \Phi_l \; .
\label{csm9}
\end{eqnarray}
The  $\Phi_k$ are real, while the  $\tilde \Phi_k$ are complex
and energy dependent in the overlapping regime. The
coefficients $b_{kl}$ and   the $\tilde \gamma_k^c$ 
are complex and energy dependent, too. 
In this regime, the differences between ${\cal H}$ and
$H$ can therefore not be simulated in a simple manner. Even
the positions of the peaks in the cross section do, generally,
not appear at the energies $E_k$ when the resonance states overlap
\cite{rep}. They are the result of interferences between the resonance
states.

It should be underlined here that  the expression (\ref{csm3}) for 
the resonance reaction part of the $S$ matrix
is  derived from the Schr\"odinger equation  (\ref{csm1})
by rewriting it in a consistent manner. Here, the eigenvalues 
$\tilde {\cal E}_k = \tilde E_k - \frac{i}{2}\,\tilde\Gamma_k$
of the effective Hamiltonian ${\cal H}$, Eq. (\ref{csm4}),
as well as the coupling matrix elements $\tilde\gamma_k^c$ 
are energy dependent functions, and the unitarity of the $S$ matrix
is garanteed.

Furthermore, the 
different  $\tilde \Phi_k {\mbox{\footnotesize $(E\!\!=\!\!E_k)$}}
$ are neither strictly orthogonal nor bi-orthogonal since 
the bi-orthogonality relation (\ref{csm5})
holds only when the  energies of both states $k$ and $l$ are equal. 
The spectroscopic studies on resonance states are performed therefore with the
wave functions being only approximately bi-orthogonal. The deviations 
from the bi-orthogonality
relation (\ref{csm5}) are, however, small as a rule. 
This drawback of the spectroscopic studies of resonance states has to be
contrasted with  the  advantage it has for the study of observable values: 
the $S$ matrix and therefore the cross section are calculated with the
resonance wave functions being strictly bi-orthogonal at every energy $E$ of
the system. Furthermore, the full energy dependence of $\tilde E_k, \tilde
\Gamma_k$ and, above all, of the coupling matrix elements $\tilde \gamma_k^c$
is taken into account in the $S$ matrix and therefore in all calculations for
observable values.

\section{Unitarity of the $S$ matrix}

The Breit Wigner one-level formula for nuclear reactions describes the 
reaction cross section with isolated resonances. The $S$ matrix 
elements for this  case   read
\begin{eqnarray}
S_{cc'} = 1 - \, i \, \sum_k \, \frac{\TW_k^{cc'}}{E-\TE_k + \,
\frac{i}{2}  \, \tilde\Gamma_k}
\label{brwi1}
\end{eqnarray} 
where $\TE_k$ and $\tilde\Gamma_k$ are the energies and widths of the 
resonance states $k$ and  $\TW_k^{cc'} \equiv \tilde\gamma_k^c \Tg_k^{c'}$.
The  $\tilde\gamma_k^c$ are  the 
partial  widths of the states $k$ relative to the channel $c$. 
The values $\TE_k, \, \tilde\Gamma_k$ and $\tilde\gamma_k^c$ 
are numbers characterizing the properties
of the resonance states $k$. Since they are energy-independent values,
the decay follows an exponential law.

For an isolated resonance state $k=1$ coupled to one channel, 
\begin{eqnarray}
S = 1 -  \, i \; \frac{\TW_1}{E-\TE_1 + \,
\frac{i}{2}  \, \tilde\Gamma_1}
\label{brwi1a}
\end{eqnarray}
in the energy range $\TE_1 - \frac{1}{2}  \, \tilde\Gamma_1 \le E \le \TE_1 + 
\frac{1}{2}  \, \tilde\Gamma_1 $, and 
$ \TW_1 = \tilde\Gamma_1$ due to the unitarity of the $S$ matrix.
The last relation follows immediately from (\ref{brwi1a}) that, 
in the one-resonance-one-channel case,  can be written as
\begin{eqnarray} 
S&=&
 \frac{E-\TE_1- \, \frac{i}{2}\, \tilde\Gamma_1}{E-\TE_1+ \, \frac{i}{2} \, 
\tilde\Gamma_1} 
\label{brwi2}
\end{eqnarray}
when $\TW_1=\tilde\Gamma_1$. The $S$ matrix (\ref{brwi2}) is unitary.

Let us now consider the unitary representation of the $S$ matrix in the 
one-channel case with two resonance states,
\begin{eqnarray}
S&=& \; \frac{(E-\TE_1- \, \frac{i}{2} \, \tilde\Gamma_1)}{(E-\TE_1 + 
\, \frac{i}{2} \, \tilde\Gamma_1)} \cdot 
\frac{(E-\TE_2- \, \frac{i}{2} \, \tilde\Gamma_2)}{(E-\TE_2 + 
\, \frac{i}{2} \, \tilde\Gamma_2)} \; .
\label{brwi2a}
\end{eqnarray}
From this expression,  a possible form of the
pole representation of the $S$ matrix can be derived,
\begin{eqnarray}
S &=& \;   1- \frac{i \, \tilde\Gamma_1}{E-\TE_1+ \, \frac{i}{2} \, 
\tilde\Gamma_1}
- \frac{i\,\tilde\Gamma_2}{E-\TE_2 + \, \frac{i}{2} \, \tilde\Gamma_2} -
\frac{\tilde\Gamma_1 \tilde\Gamma_2}{(E-\TE_1+\frac{i}{2} \tilde\Gamma_1)
(E-\TE_2+\frac{i}{2} \tilde\Gamma_2) }
\nonumber \\
&=& \; 1- \frac{1}{E-\TE_1+ \, \frac{i}{2} \, \tilde\Gamma_1}\; 
\Bigg( i\, \tilde\Gamma_1 + 
\frac{\tilde\Gamma_1 \tilde\Gamma_2}{2E-(\TE_1 + \TE_2) + 
\, \frac{i}{2} \, (\tilde\Gamma_1 + \tilde \Gamma_2)}\Bigg) 
\nonumber \\
& & \; \; -
\frac{1}{E-\TE_2+ \, \frac{i}{2} \, \tilde\Gamma_2}\; \Bigg( i\, 
\tilde\Gamma_2 + 
\frac{\tilde\Gamma_1 \tilde\Gamma_2}{2E- (\TE_1+ \TE_2) + 
\, \frac{i}{2} \, (\tilde\Gamma_1 + \tilde \Gamma_2)}\Bigg)
\; .
\label{brwi3}
\end{eqnarray}
It follows 
\begin{eqnarray}
S = 1 \, - \, i \, \sum_{k=1,2} \frac{\TW_k}{E-\TE_k + 
\, \frac{i}{2} \, \tilde\Gamma_k}
\label{brwi4}
\end{eqnarray}
in complete analogy to (\ref{brwi1}), with 
\begin{eqnarray}
\TW_k & = &\tilde \Gamma_k \; \Bigg( 1 - \,i \, 
\frac{\tilde\Gamma_{l}}{2E-(\TE_k + \TE_{l}) +
\, \frac{i}{2} \, (\tilde\Gamma_k +\tilde \Gamma_{l})} \; \Bigg)
\label{brwi5}
\end{eqnarray}
and $k,l = 1,2$, ~$l \ne k$.
These equations show that the coupling coefficients $\TW_k$ are complex and 
energy dependent,  that $\TW_k$ has a resonance 
behaviour at the energy $(\TE_k + \TE_l)/2$ with the width 
$(\TG_k + \TG_l)/2$, and that the
energy dependence of the two values $\TW_k$ and $\tilde\Gamma_k$ is different
in the overlapping regime.
In the energy region of the  resonance behaviour of $\TW_k$
caused by a neighboured resonance state $l$, the $S$ matrix contains terms 
being nonlinear in energy.

When the widths of the two states are equal, $\TW_k\to 0$ with $E\to 
(\TE_k + \TE_l)/2$.
At large distance $E \gg \TE_1 + \TE_2$,  follows $\TW_k \to 
\tilde\Gamma_k $. 
In this case, the two resonance states behave as isolated ones.
When the positions $\TE_k, \TE_l$ of the two resonance states are
outside of resonance region of $\TW_k$ and $\TW_l$, they can
also be considered, to a good approximation, as
isolated, and $\TW_k \approx \tilde\Gamma_k, \; \TW_l \approx \tilde\Gamma_l$.

The nonlinear term creates some deviation in the resonance line shape
from the linear Breit-Wigner one.
This can be seen best in the case when the $S$ matrix has a 
double pole, i.e. $\TE_1 = \TE_2 \equiv \TE_d$ and $\tilde\Gamma_1 
=\tilde \Gamma_2 \equiv 
\tilde\Gamma_d$. In such a case, the $S$ matrix (\ref{brwi4}) reads
\begin{eqnarray} 
S & = & 1 - 2\, i \, \frac{\TW_d}{E-\TE_d + \, \frac{i}{2} \,\tilde \Gamma_d}
\nonumber \\
&=& 1 - 2\, i \, \frac{\tilde\Gamma_d}{E-\TE_d + \, \frac{i}{2} \, 
\tilde\Gamma_d} -
\frac{\tilde\Gamma_d^2}{(E-\TE_d + \, \frac{i}{2} \, \tilde\Gamma_d)^2}
\; .
\label{brwi6}
\end{eqnarray}
The second term corresponds to the usual linear term while the third term is 
quadratic (see  \cite{newton}).  
The interference between these two parts has been
illustrated in \cite{mudiisro}, where the cross section is shown for the case 
of two resonance states coupled to one channel. The energies and widths
of the two resonance states  are the same, creating a double 
pole of the $S$ matrix.
The asymmetry of the line shape of both peaks in the cross section agrees with
(\ref{brwi6}).  A similar picture has been obtained 
in, e.g., laser induced continuum structures in atoms 
with a double pole of the $S$ matrix \cite{kylstra,marost}, in  
atom-surface collisions \cite{bosanac}, transmission in quantum
scattering systems \cite{buttiker}, in a double barrier potential \cite{mond},
a double-square-well system \cite{vanroose} and in a toy model for  
the conductance through a small quantum dot \cite{silva}. 

According to (\ref{brwi5}), the asymmetry  of narrow resonances is 
usually larger than
that of broad resonances:  when $\tilde\Gamma_1 \gg \tilde\Gamma_2$,
it follows $~\TW_1 \approx \tilde\Gamma_1$ 
while the corrections from $\tilde\Gamma_1$ to $\TW_2$ 
can mostly not be neglected.
In any case, the nonlinear  term in $\TW_k$, eq. ({\ref{brwi5}), 
causes a non-exponential decay of the two resonance states. 
Only when the line shape of a certain resonance $k$ is of Breit-Wigner
type and $\tilde\Gamma_k$ is almost constant in a large energy region
around $\TE_k$, the states $k$ will decay according to an exponential law
\cite{newton}. 
 
When the two resonance states lie at the same energy, but their
widths are different, then follows
from (\ref{brwi4}) and (\ref{brwi5}) that
the contributions from both resonance states 
annihilate each other  at $E=\TE_1=\TE_2$, i.e.
the cross section vanishes  at that energy where the two resonance
states lie. This
destructive interference has been traced numerically in \cite{mudiisro} 
for two resonance states coupled to one channel 
by varying the coupling strength
between the states and the continuum. When $\tilde\Gamma_1 \gg 
\tilde\Gamma_2$, the narrow
resonance appears as a dip in the cross section that is determined mainly by
the broad resonance. This is in accordance with (\ref{brwi5}): for $E\to 
\frac{1}{2} (\TE_1+\TE_2)$, follows $\TW_1 \to \tilde \Gamma_1$ and 
$\TW_2 \to - \tilde\Gamma_2$.

For illustration, the cross section calculated with two neighbouring resonance
states is shown in Fig. \ref{fig1}. The width of one of the states is 
fixed to $\TG_1=0.05$ (in arbitrary units) while that of the other one 
is varied between $\TG_2 = 0.01$ and 5.0. When $\TG_2 \gg \TG_1$, it is
$\TG_2 \approx$ const in the energy region of the narrow resonance, and the
cross section shows a dip at the energy $\TE_1$. In this case, the broad
resonance plays the role of a background for the narrow resonance. 
When the widths of both states are equal, the structure of the cross section 
is similar to that caused by (\ref{brwi6}) at the double pole of the 
$S$ matrix. When $\TG_2 < \TG_1$, the two peaks in the cross section are no
longer symmetrical in relation to  the center at $E=8$. 

The corresponding coupling coefficients $\TW_1$ and $\TW_2$ (Figs.
\ref{fig2} and \ref{fig3})
show a resonance-like behaviour at $E=(\TE_1 + \TE_2)/2$. 
While the absolute values
show the same tendency for both states, their phases  behave differently. 
The phase of the broad state is almost not influenced 
by the interaction with the narrow one, while the phase of the narrow state 
jumps by $2\pi$ at the energy $E=(\TE_1 + \TE_2)/2$. 
This is caused by the minimum  of Re$(\TW_k)$ that is
reached at $E = (\TE_k + \TE_l)/2$. The minimum value is 
$ \TG_k - \TG_l \approx \TG_k$ 
for the broad state, but $- \TG_l$ for the narrow one.
In both cases, Im$(\TW_{k(l)})$ oscillates and vanishes at
$E = (\TE_k + \TE_l)/2$.
These results are  in agreement with the fact that the 
broad resonance state plays the role of a background for the narrow resonance
state. 

Of special interest is  the case  $\TG_1 = \TG_2$. The resonance
behaviour of the coupling coefficients appearing at  $E=(\TE_1 + \TE_2)/2$,
is independently of the distance  $|\TE_1 - \TE_2|$ 
of the two resonance states. It reflects
the properties of a double pole of the $S$ matrix. 
One clearly sees the phase jump by $\pi$ at  $E=(\TE_1 + \TE_2)/2$
(Figs. \ref{fig2} and \ref{fig3}).
This energy, $E^{\rm cr}\equiv (\TE_1 + \TE_2)/2$, is the critical value at 
which the wave functions of the two states are exchanged when the energy is 
parametrically varied (see Section IV). Here, $\TW=0$.
However, the phase jump of $\pi$ does not appear at every zero of 
the cross section, see Figs. \ref{fig1} to \ref{fig3}. 
It follows further that the
resonance behaviour of the coupling coefficients plays a role only for
resonance states lying near to one another. 

As can be seen in Fig. \ref{fig1}, the
interferences between the different resonance states cause, in the
one-channel case, a separation of the peaks in the cross section. 
In this manner, the interferences may feign the 
existence of well isolated resonance states in spite of their strong 
overlapping. An extreme case are the two separated peaks appearing 
in the cross section with a double pole of the $S$ matrix 
\cite{rep,marost4a,mudiisro} or in its neighbourhood, Fig. \ref{fig1}.

The line shape of the peaks in the cross section
is described usually by means of energy independent Fano parameters. 
A recent example are the experimentally observed narrow peaks 
in the conductance 
through a quantum dot controlled by varying the strength of
the magnetic field  \cite{kob}.
The energy independent Fano parameters are related to
a representation of the $S$ matrix (\ref{brwi4}) 
with energy independent $W_k'$ [instead of the energy dependent $\TW_k$
in (\ref{brwi5})]:
\begin{eqnarray}
W'_k & = & \tilde\Gamma_k \; \Bigg( 1 - \,i \, 
\frac{\tilde\Gamma_{l}}{\TE_k - \TE_{l} -
\, \frac{i}{2} \, (\tilde\Gamma_k - \tilde\Gamma_{l})} \; \Bigg) \; .
\label{brwi5a}
\end{eqnarray}
In the literature, mostly the  $W'_k$ are used since they provide 
the standard parametrization of the $S$ matrix. A recent example 
is the analysis of the bi-orthogonality of resonance wave functions in the
molecule NO$_2$ \cite{greben}. The
representation (\ref{brwi5a}}) is equivalent to (\ref{brwi5}), 
except in approaching a double pole of the $S$ matrix where (\ref{brwi5a}) 
has a singularity in contrast to   (\ref{brwi5}). 
The $W_k'$ lose, however, their physical meaning as coupling matrix 
elements between resonance states and continuum in the
overlapping regime \cite{marost4a}.  For this reason, we will 
not consider them in this paper.  Instead, we will use the $\TW_k$
that are meaningful also around double poles of the $S$ matrix
according to (\ref{brwi6}).

It is easy to generalize the study to more than two resonance
states. Suppose
\begin{eqnarray}
S&=& \; \prod_{n=1}^N \, \frac{E-\TE_n- \, \frac{i}{2} \, 
\tilde\Gamma_n}{E-\TE_n + \, \frac{i}{2} \, \tilde\Gamma_n}
\label{more1}
\end{eqnarray}
instead of (\ref{brwi2a}). In analogy with (\ref{brwi4}) and (\ref{brwi5}),
it follows
\begin{eqnarray}
S = 1 \, - \, i \, \sum_n \frac{\TW_n}{E-\TE_n + \, \frac{i}{2} 
\,\tilde \Gamma_n}\, 
\label{more4}
\end{eqnarray}
with 
\begin{eqnarray}
\TW_n & = &\tilde \Gamma_n \; \Bigg( 1 - \,i \,\sum 
\frac{\tilde\Gamma_m}{X_n + X_m} -
\, \sum \frac{\tilde\Gamma_m \cdot 
\tilde\Gamma_l}{X_nX_m + X_nX_l + X_mX_l} \;
- ... \Bigg)
\label{more5}
\end{eqnarray}
and $X_n \equiv E-\TE_n+\frac{i}{2}\tilde \Gamma_n$. 
The sum in the second term of
(\ref{more5}) is running over $m \ne n$ and that in the third term over 
$m\ne n $ and $l\ne m,n$. The denominator $X_n+X_m$ of the second term
is linear in $E$ while that of the third term, 
$X_nX_m+X_nX_l+X_mX_l$, is quadratic in $E$. In any case, the $\TW_n $ 
depend on energy in a non-trivial manner.

As follows from (\ref{more4}) and  (\ref{more5}), the coupling
coefficients at a triple pole of the $S$ matrix are $\TW_n=
\tilde\Gamma_n /3 $ (in
contrast to $\TW_n=0$ at a double pole). Here, the $S$ matrix contains terms 
up to third order.
When one of the widths is much larger than the other ones in the
three-resonance case, $\tilde\Gamma_l \gg
\tilde\Gamma_k$ ~($k=n,m$),  it is $\TW_l \approx \tilde\Gamma_l$ and 
$\TW_{k} = -\tilde \Gamma_{k}$ when $E = \frac{1}{2}(\TE_l+\TE_{k})$.
These relations are in complete analogy with 
those obtained for the two-resonance case.

In Figs. \ref{fig4} and \ref{fig5}, the cross sections 
with three resonance states 
are shown, two of which are lying symmetrically around the position of
the third one at $E=\TE_3=8$,
as well as the coupling coefficients $\TW_3$. 
The different curves in Fig. \ref{fig4} are
obtained by varying the width $\TG_3$ between 0.05 and 5. In the first case,
the widths of all three states are equal while in the other cases, the middle
state overlaps the narrower ones. When $\TG_3 \gg \TG_k$, ~($k=1,2$), 
the broad state can be considered as a ''background'' for the two other
ones: they appear as dips in the cross section. This can be seen  better 
in Fig. \ref{fig6} where one broad resonance state ($\TG_3= 3.0$) is shown
that overlaps  two narrow ones ($\TG_1=\TG_2=0.05$). 
In both cases, a peak appears in the middle of the spectrum 
at $E=8$ in contrast to the two-resonance case Fig. \ref{fig1}.
In Fig. \ref{fig5}, the widths
of all three states are kept constant while the distance between them is
varied. Altough the peaks in the cross section seem to be well isolated from
one another for $(\TE_{k} - \TE_{k\pm 1})/(\TG_k+\TG_{k\pm 1}) = 2.5$ (full
curve),  the interferences between the resonance states are not vanishing.

Interesting are again the phase jumps of the coupling coefficients $\TW_3$
at the critical 
values of the energy $E^{\rm cr}=(\TE_k + \TE_3)/2; ~k=1,2$.
They are mostly smaller than those of $\TW_1$ and $\TW_2$  (not shown).
The influence of a broad resonance state on the coupling coefficients
of narrow ones  is illustrated in  Figs. \ref{fig6} and \ref{fig7}
with the state 3 being, respectively, much broader and much narrower than
the states 1 and 2. At the energies 6 and 9,
the states 1 and 2, respectively, interact with the state 3 while  the
interference picture at the energy 7 is caused by the two states 1 and 2
with equal widths. The phase jump of $\pi$ at this energy (Fig. \ref{fig7}) 
is reduced in the presence  of the broad resonance state (Fig. \ref{fig6}).   
Also  $|\TW_1|$ and $|\TW_2|$ differ in the two cases without and with a
broad state.

Measurements of phase and magnitude of the reflection and transmission
coefficients of a quantum dot are performed in \cite{heiblum}. As a
result, the phases of the dot's transmission and reflection
coefficients change abruptly by
about $\pi$ at some energy in the resonance peak. The phase changes 
are very similar to those observed in the present calculations
for the $\TW$ in the one-channel case (see the figures). As discussed above,
they are related to the unitarity of the $S$ matrix 
in the overlapping regime.
The results are expected to be similar for the case with more
channels (or terminals)  since, as will be shown in Sect. IV, they are 
characteristic of the  intrinsic wave functions $\TP_k$ and $\TP_k^{\rm ch}$,
respectively, of the system. 

In the consideration presented in the present section, the $\TE_k$ and $\TG_k$
are assumed to be independent of the energy. This is, mostly, a good
approximation in the energy range of the resonance state $k$. In any case,  
the energy dependence of the coupling coefficients $\TW_k$ arises 
primarily from
their resonance behaviour caused by a neighbouring resonance state,
Eq. (\ref{brwi5}). It may be
influenced, of course, by the energy dependence of $\TE_k$ and $\TG_k$, 
especially when the levels repel or attract each other by 
varying a certain parameter.

\section{Wave functions near avoided level crossings in the complex plane}

The coupling coefficients between system and 
continuum are defined by  $(\tilde\gamma_k^c)^2$,  
Eq. (\ref{over1}).  Their energy dependence and phase are determined,
in the one-channel case, by the energy dependence  
and phase of the $(\tilde \Phi_k)^2$
(after removing the common phase and energy dependence 
caused by the $\xi_E^c$). Much can, therefore, be learned 
on the behaviour of
the coupling coefficients between system and environment
from a study of the wave functions $\TP_k$.  

Characteristic of the overlapping regime are avoided level crossings in the
complex plane.  
At an avoided level crossing, the wave
functions of the two crossing states are exchanged. This fact is very well
known for a long time for discrete states (Landau-Zener effect). It holds
also for resonance states in the adiabatic limit
\cite{rep}. The difference
between avoided crossings of discrete and resonance states consists mainly 
in the fact that
resonance states may cross in the complex plane even when the interaction
between them is non-vanishing and, furthermore, that the crossing may lead 
not only to level repulsion but also to level attraction. While level
repulsion is accompanied by the tendency to equilibrate the widths of the
resonance states, level attraction is accompanied by the formation of
different time scales in the system. 

The avoided level crossing of resonance states 
can be traced back to a branch point in the complex plane \cite{ro01}. 
When the conditions for crossing of the two levels are fulfilled,
the branch point in the complex plane is nothing else than a double pole of
the $S$ matrix (see also \cite{mois}). 
In any case, the wave functions of the two levels are
exchanged at the critical value of the parameter at
which the two levels avoid  crossing (or cross in one point). Here, 
either the widths or
the energies (positions) of both states are equal. In the first case, the 
avoided crossing happens, as for discrete states, 
in the energies of the resonance states traced as a function of the 
considered parameter. In the second case, however, it appears in their widths.

Further studies have shown that the wave functions at the critical value 
$a^{\rm cr}$ of the parameter $a$
are exchanged according to 
\begin{eqnarray}
\tilde \Phi_k \to \, \pm \, i \, \tilde \Phi_l
\label{wf1}
\end{eqnarray}
where $k$ and $l $ are the two crossing states. 
This relation has been obtained
analytically \cite{ro01} as well as in a numerical study of atoms in a laser
field \cite{marost}. It is related to nonlinear terms that appear in the
Schr\"odinger equation due to the
biorthogonality of the eigenfunctions of the effective Hamiltonian
${\CH}$. At $a^{\rm cr}$, 
the sign of the imaginary part of the wave function $\tilde \Phi_k $
jumps from + to $-$ (or opposite)
even when the two states avoid crossing and $A_k \equiv |\TP_k|$ 
remains finite \cite{ro01}.
That means, in a certain parameter range  $a^{\rm min} \le a^{\rm cr} 
\le a^{\rm max}$, the  wave functions of the two states 
$k$ and $l\ne k$ are mixed, 
\begin{eqnarray}
\tilde \Phi^{\rm ch}_k & = & \beta_k \, \tilde\Phi_k \pm i \, \beta_{l} \, 
\tilde\Phi_{l} \, .
\label{wf2}
\end{eqnarray}
The wave functions $\tilde \Phi_k^{\rm ch}$ 
change  smoothly (without any jump of the sign of their components)
\begin{eqnarray}
{\rm from}\quad \; \; \beta_k \to \pm 1, \, \beta_l\to 0 \quad & 
{\rm at} & \quad 
a \to a^{\rm min} < a^{\rm cr} 
\nonumber \\
{\rm  to }\qquad \; \,
\beta_k \to 0, \, \beta_l \to \pm 1 \quad & {\rm  at} & \quad  a 
\to  a^{\rm max} > a^{\rm cr} 
\; .
\label{wf3}
\end{eqnarray}
The values $a^{\rm min}$ and $a^{\rm max}$
may be quite different  from one another \cite{ro01}. 
Only in the case the avoided level
crossing shrinks to one point, being the double pole of the $S$ matrix,  
$\beta_{k}=0 $ or $\pm 1$  for all $a$ but $a^{\rm cr} $. 
In any case, the two wave functions  $
\frac{1}{\sqrt{2}} \, (\tilde\Phi_k \pm i \, \tilde\Phi_{l})$
and $\frac{1}{\sqrt{2}} \, 
(\tilde\Phi_l \mp i \, \tilde\Phi_{k}) $
remain unchanged at $a=a^{\rm cr} $ under the exchange  (\ref{wf1}).

Hence, in the parameter range $ a^{\rm min} < a < a^{\rm max}$,
the wave functions of the two states are $\tilde\Phi_k^{\rm ch}$,
but not $\tilde\Phi_k$ ~($k=1,2$). 
The two wave functions are restored, after the exchange at  $a^{\rm cr}$,
only at  $a \ge a^{\rm max}$. In other words, using the representation
\begin{eqnarray}
\tilde \Phi_k^{\rm ch} = |\tilde \Phi_k^{\rm ch}| \; e^{i\, \theta_k}
\; ,
\label{wf4}
\end{eqnarray}
$\theta_k$ depends on $a$ when $ a^{\rm min} < a < a^{\rm max}$. 
Beyond this parameter area, $\theta_k$ depends on $a$  
much weaker (if at all). After removing a common phase factor, 
it follows from (\ref{wf3})
for an avoided level crossing near the double pole:
$\theta_k \to \pm \pi/4$ and $\pm 3\pi/4$, respectively,
in approaching   $a^{\rm cr}$ and $\theta_k \to 0$ or $\pi$ in 
approaching $ a^{\rm min}$ or  $ a^{\rm max}$.
At $a^{\rm cr}$  the states are chiral.  

The comparison of (\ref{wf1})  with experimental data obtained from
microwave cavities
has been discussed in detail in \cite{ro02}. All the data published in
\cite{lauber,demb1} for the case of two levels
of the system that are well isolated 
from the other ones, can be explained by means of (\ref{wf1}).
Some  chirality appears: left and right turns around 
the double pole of the $S$ matrix with $\beta_k = \beta_l$
are different from one another according to (\ref{wf2}) and
(\ref{wf3}).

Phase changes of the wave functions in the conductance through 
a microwave cavity have been considered  in   \cite{seha,ishio}.
The question is studied to what extent the transport through the cavity
changes the structure of its internal wave functions. 
It is demonstrated theoretically as well as experimentally \cite{seha} that 
the standing waves of the original cavity are transformed more or less
completely into running waves propagating from the entrance antenna to 
different exit ports. This is expressed by the two  limiting
cases: (i) the real and imaginary parts of the wave functions are 
strongly correlated,  and (ii) they are completely uncorrelated. 
In the last case, the real and imaginary parts of the wave functions  
of the resonance states   evolve independently
in the open microwave cavity.

In \cite{demb2}, the phase difference between two modes has been measured
in a cavity composed of two almost identical semi-circular parts. The 
two modes are each localized in one of the semi-circular parts of the cavity
and are excited separately by appropriately positioned dipole antennas.
The corresponding two  eigenvalues are well separated from all the other ones. 
By varying two parameters (designed here together by $a$) of the cavity, 
their avoided crossing in the 
complex plane to an (almost) true crossing can be traced. Furthermore, 
the eigenfunctions are
studied by mapping the distributions of the electric field.  
Finally, the phase difference 
$\Delta = \theta_k - \theta_l$ between the antennas 
has been found for different distances of $a$ from the critical 
value $a^{\rm cr}$.    
The results obtained in the experiment \cite{demb2}
are $\Delta \to \pi/2$ for $a \to a^{\rm cr}$
and  $\Delta = \pi$ for $a$ beyond the range  $ a^{\rm min} < a < a^{\rm
max}$. They agree with   (\ref{wf2}) and   (\ref{wf3})
by using the representation (\ref{wf4}) \cite{comment}.

According to ({\ref{over1}), the
phase  of the coupling coefficients  $(\tilde\gamma_k^c)^2$ 
is determined by that  of  $(\TP_k)^2$. Considered as a function of a certain
parameter, the phases of both expressions vary,
in the one-channel case,  in the same manner. Since
the parameter may be also the energy of the system \cite{rep},  
the phase of the  $(\tilde\gamma_k^c)^2$ varies generically
with the energy in the same manner as 
the phase of the $(\TP_k)^2$. The  results discussed above are in full 
agreement  with those following immediately from the unitarity of the $S$
matrix, Figs. \ref{fig2} to \ref{fig7}.
Even the jumps by $\pi$ appearing in the phases of the coupling coefficients  
$\TW$ in Figs. \ref{fig2}, \ref{fig3} and \ref{fig7}
can be explained by Eqs. (\ref{wf1}) and   (\ref{wf2}).
In other words, Eqs. (\ref{wf1}) to (\ref{wf3})
coincide with the postulation of the  unitarity of the $S$ matrix.
In any case, nonlinearities are responsible for the 
energy dependence of the coupling coefficients between system and continuum.

In realistic systems, mostly  more than two resonance states 
are coming close
to one another, i.e. avoided crossings of more than two resonance states 
take place
in the parameter range   $ a^{\rm min} < a < a^{\rm max}$. 
As an extreme case, the  branch point in the complex plane may be a
multiple pole of the $S$ matrix. The relation between the different 
wave functions
is, in such a case, more complicated than that for two states   where  
the wave functions of only two resonance states are exchanged
according to (\ref{wf1}), mixed in
the range $ a^{\rm min} < a < a^{\rm max}$ and restored beyond this range.
Such a situation is studied experimentally \cite{lauber} as well as
theoretically in different approaches \cite{jung,ro01,pisto,mois1}, compare 
also Figs. \ref{fig2} and \ref{fig3} for two resonance states with
Figs. \ref{fig4} and \ref{fig5} for three resonance states as well as 
Fig. \ref{fig6} with Fig. \ref{fig7}.

\section{Concluding remarks}

The resonance phenomena are described well by 
two ingredients also at high level density. The first ingredient is
the effective Hamiltonian ${\cal H}$ that contains all the  basic
structure information involved in the Hamiltonian
$H$, i.e. in the Hamiltonian of the corresponding closed system with discrete
eigenstates.  Moreover, ${\cal H}$ contains the coupling matrix
elements $W_k^{cc'}$ between discrete states and continuum
that account for the changes of the system 
under the influence of its coupling to the continuum. These matrix
elements are responsible
for the non-Hermiticity of ${\cal H}$ and  its complex eigenvalues
which transfer the discrete states into resonance states and
determine not only their positions but also their (finite) lifetimes.

The second ingredient is the unitarity of the $S$ matrix that has to be
fulfilled in all calculations of resonance phenomena. The unitarity of the
$S$ matrix causes a non-trivial energy dependence of the coupling 
matrix elements  $\TW_k^{cc'}$
between resonance states and continuum. This energy dependence
becomes decisive in the overlapping regime even in the case 
the lifetimes of the overlapping states are very different from one another
and the different long-lived states seem to be well isolated from one another.
It is taken into account in the unified description of structure and
reaction aspects since any statistical or perturbative  
assumptions are avoided in solving 
the basic equation (\ref{csm1}). The  unitarity of the $S$
matrix influences also  the phases of the wave functions of the
resonance states that change generically in approaching avoided level 
crossings in the complex plane.

In the non-overlapping regime, both ingredients are fulfilled in almost all
theoretical approaches. Here,
the wave functions and positions of the 
resonance states are described, to a good approximation, 
by the wave functions and
positions of the discrete states of the corresponding closed system.  
The coupling matrix elements between system and continuum can be calculated
by means of the wave functions of the discrete
states. They are energy independent, to a good approximation.

In the overlapping regime however with many avoided level crossings,
the wave functions of the resonance  states suffer phase changes.
Furthermore, the coupling coefficients of the
resonance states to the continuum show a resonance-like behaviour 
caused by the interaction with a neighboured resonance state.
Both effects are described by nonlinear terms appearing in the
Schr\"odinger equation and the $S$ matrix, respectively.
They are related to one another and
can not be neglected at high level density.

As a conclusion of these results,  the coupling of a quantum
system to the environment may change its properties. 
The changes are small as long as the coupling strength between system 
and environment
is smaller than the distance between the individual states of the 
unperturbed system, i.e. smaller than the distance between the 
eigenstates of the Hamiltonian $H$.
The changes can, however, not be neglected when the coupling to the continuum 
is of the same order of magnitude as the level distance or larger. 
In such a case, the changes can be described neither by perturbation theory
nor by introducing statistical assumptions for the level distribution.
Here, nonlinear effects become important which cause 
a redistribution of the spectroscopic properties of the system and,
consequently, changes of its features.

Under the influence of the coupling to the continuum, not only level 
repulsion but also level attraction may appear that are accompanied by
the tendency to form a uniform time scale for the system in the first case,
but different time scales in the second case. 
The formation of different time scales
in an open quantum system that is accompanied by level attraction, is
accompanied also by the appearance of a non-trivial 
energy and phase dependence of the coupling coefficients $\TW$. 
The use of an effective non-Hermitian Hamilton operator in describing 
scattering processes in the overlapping regime
is therefore meaningful  only when, at the same time, the
energy dependence of the  $\TW$ is considered.

\vspace{.5cm}

\noindent
{\bf Acknowledgment:}
I am indebted to A.I. Magunov,  N. Moiseyev 
and E. Persson for valuable discussions and to 
H. Schomerus for a critical reading of the manuscript.

\begin{figure}
\begin{center}
\includegraphics[width=14cm,angle=0]{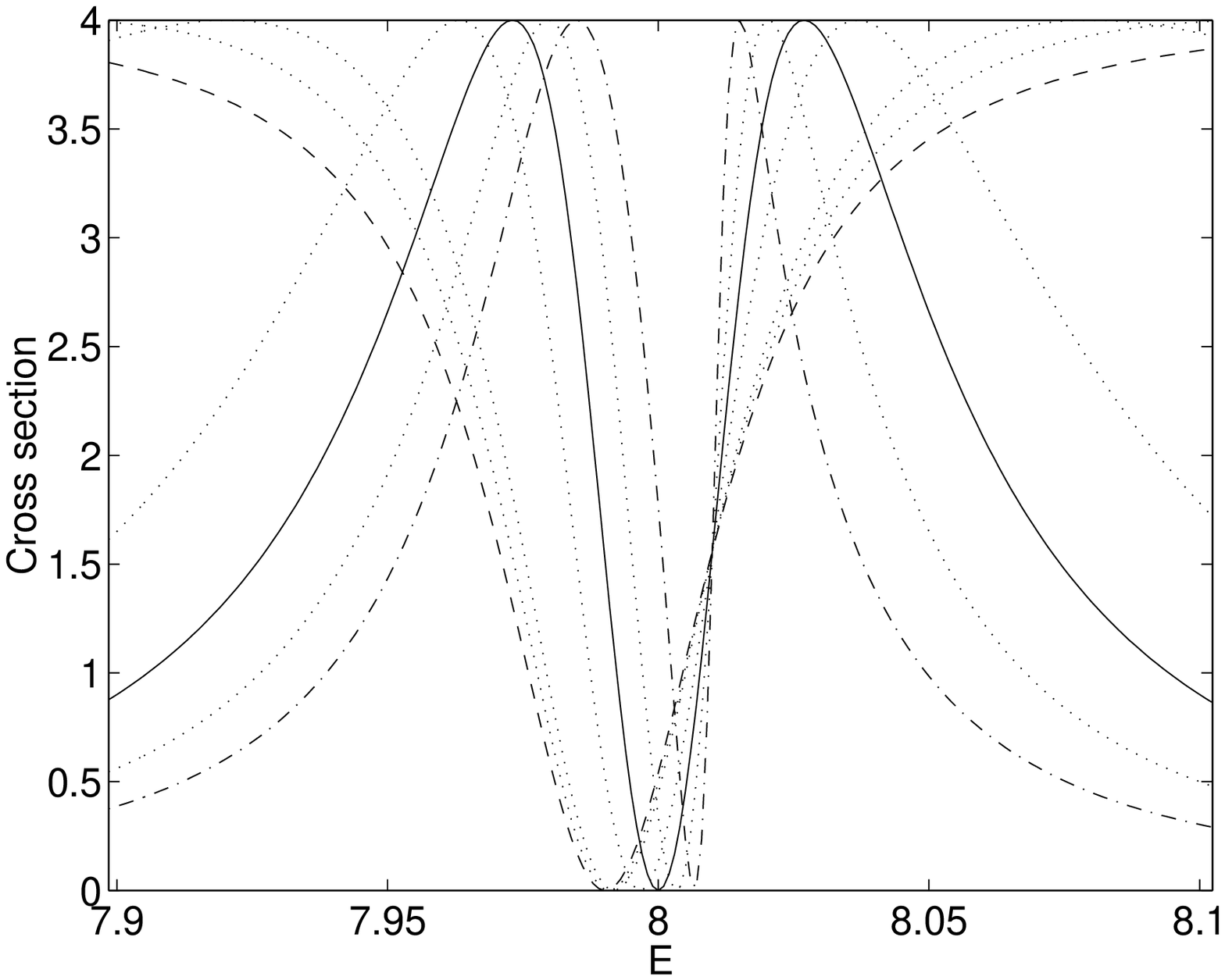}
\end{center}
\vspace*{.5cm}
\caption{
Cross section with two resonance states at $\TE_1=7.99$ and $\TE_2=8.01$.
The width of one of the states is fixed to 
$\TG_1 =  0.05$, while that of the other state is varied:
~$\TG_2=5.0$ (dashed curve),
$\TG_2=0.05$ (full curve), $\TG_2=0.01$ (dash-dotted curve).
The dotted curves are calculated with $\TG_2 = 1.0, ~0.5, ~0.1, ~0.025$,
respectively.
Cross section and energy are given in arbitrary units.
}
\label{fig1}
\end{figure}

\begin{figure}
\begin{center}
\includegraphics[width=12cm,angle=0]{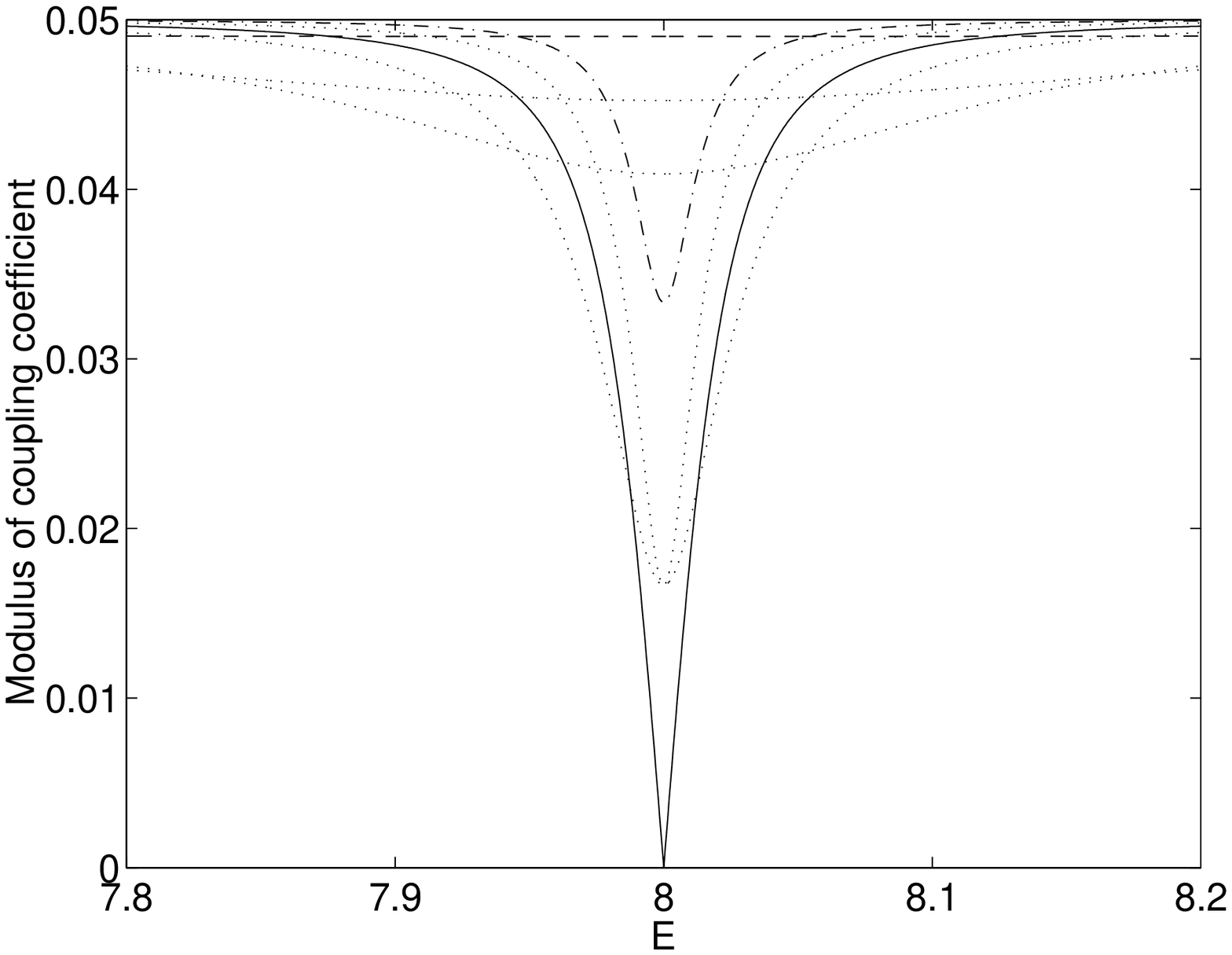}
\includegraphics[width=12cm,angle=0]{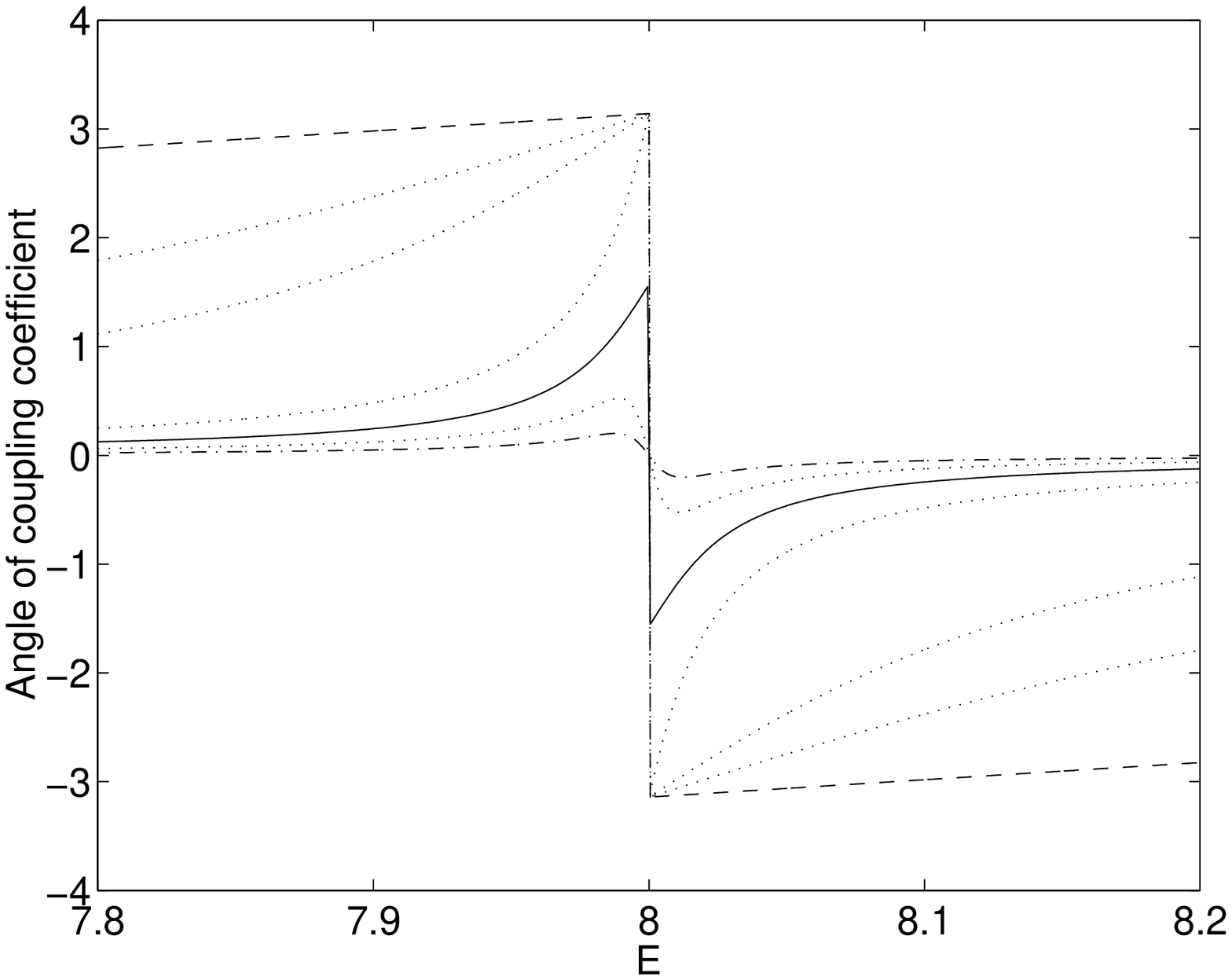}
\end{center}
\vspace*{.5cm}
\caption{
Coupling coefficient $\tilde W_1$ of the resonance state at $\TE_1=7.99$.
The width is $\TG_1=0.05$. The position and width of the other state is
as in Fig. \ref{fig1}.
}
\label{fig2}
\end{figure}

\begin{figure}
\begin{center}
\includegraphics[width=12cm,angle=0]{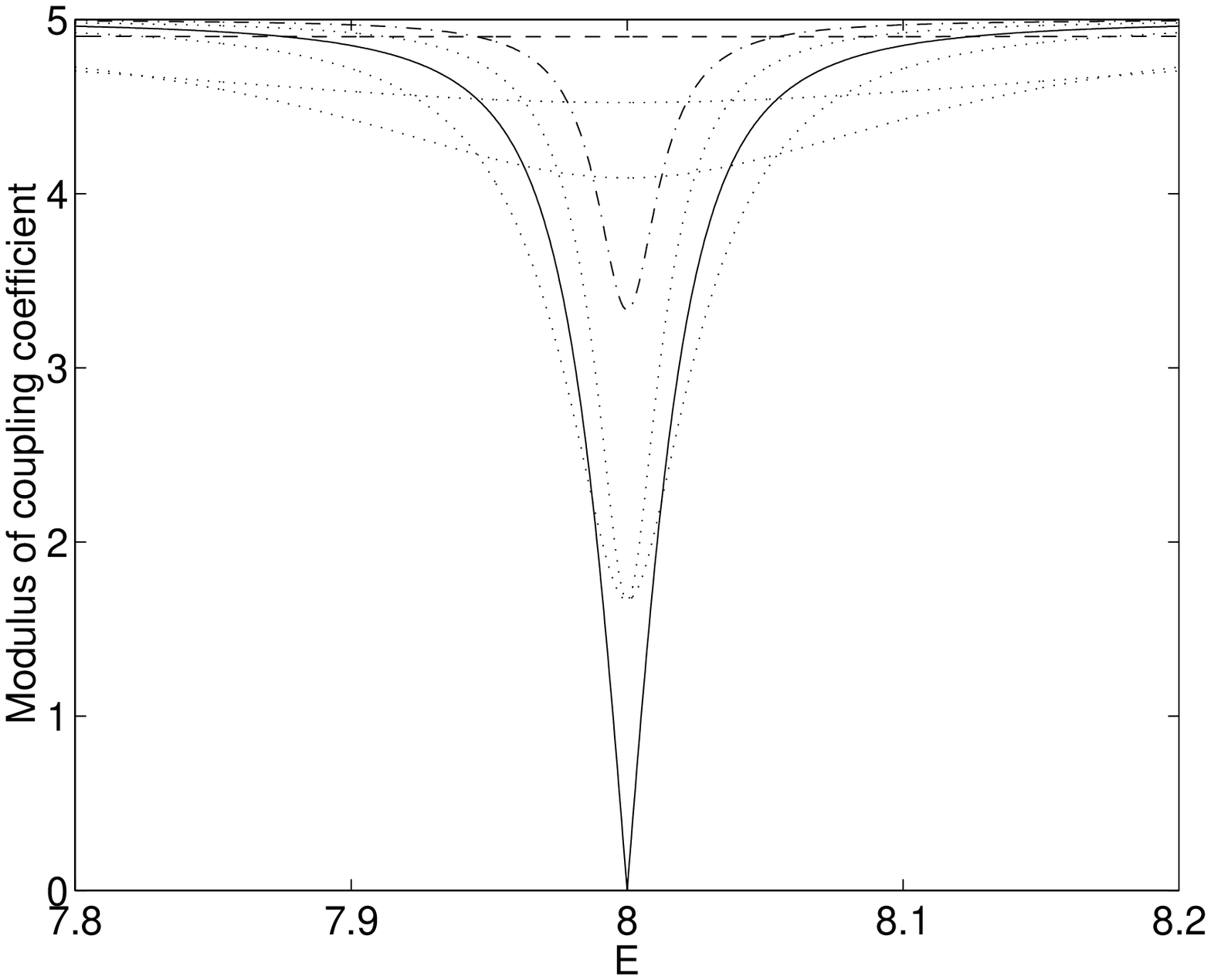}
\includegraphics[width=12cm,angle=0]{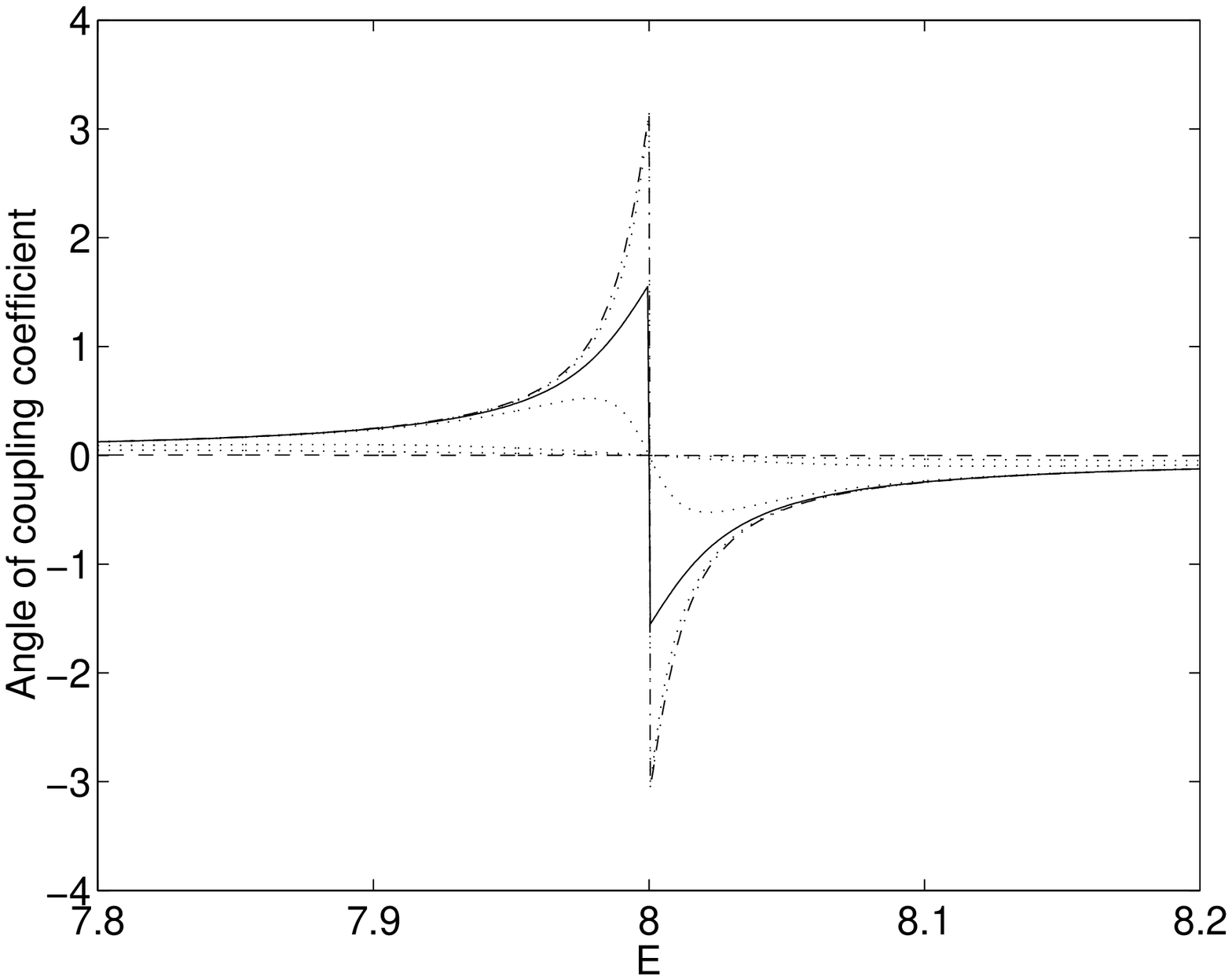}
\end{center}
\vspace*{.5cm}
\caption{
Coupling coefficient $\tilde W_2$ of the resonance state at $\TE_2=8.01$. 
The position and width of the other state as well as the width $\TG_2$ is
as in Fig. \ref{fig1}.
In the upper part of the figure, $|\TW_2|$ is multiplied by 1 (dashed curve),
100 (full curve) and 500 (dash-dotted curve) and 
by 5, 10, 50, and 200, respectively (dotted curves).
}
\label{fig3}
\end{figure}

\begin{figure}
\begin{center}
\includegraphics[width=8cm,angle=0]{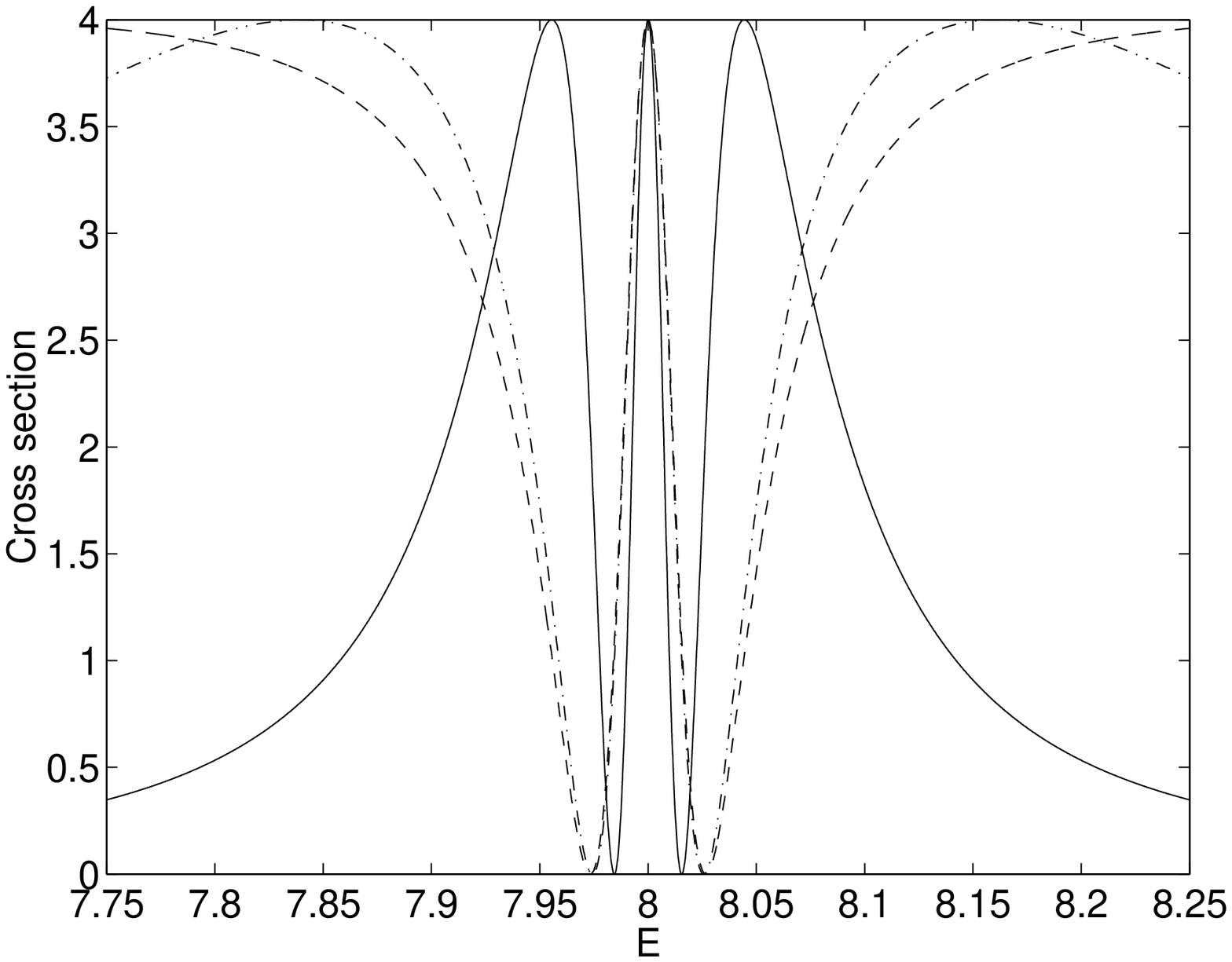}\\
\includegraphics[width=8cm,angle=0]{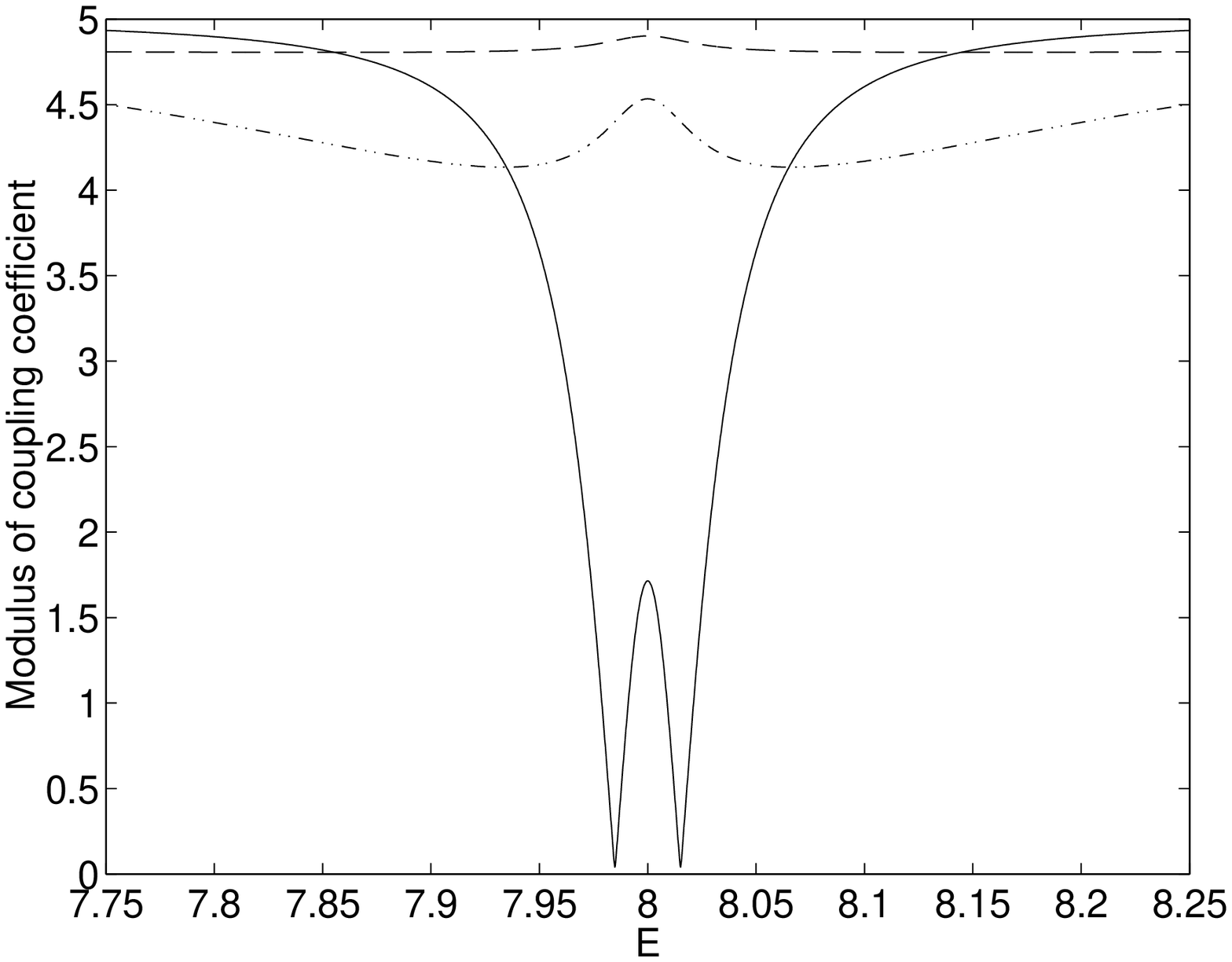}\\
\includegraphics[width=8cm,angle=0]{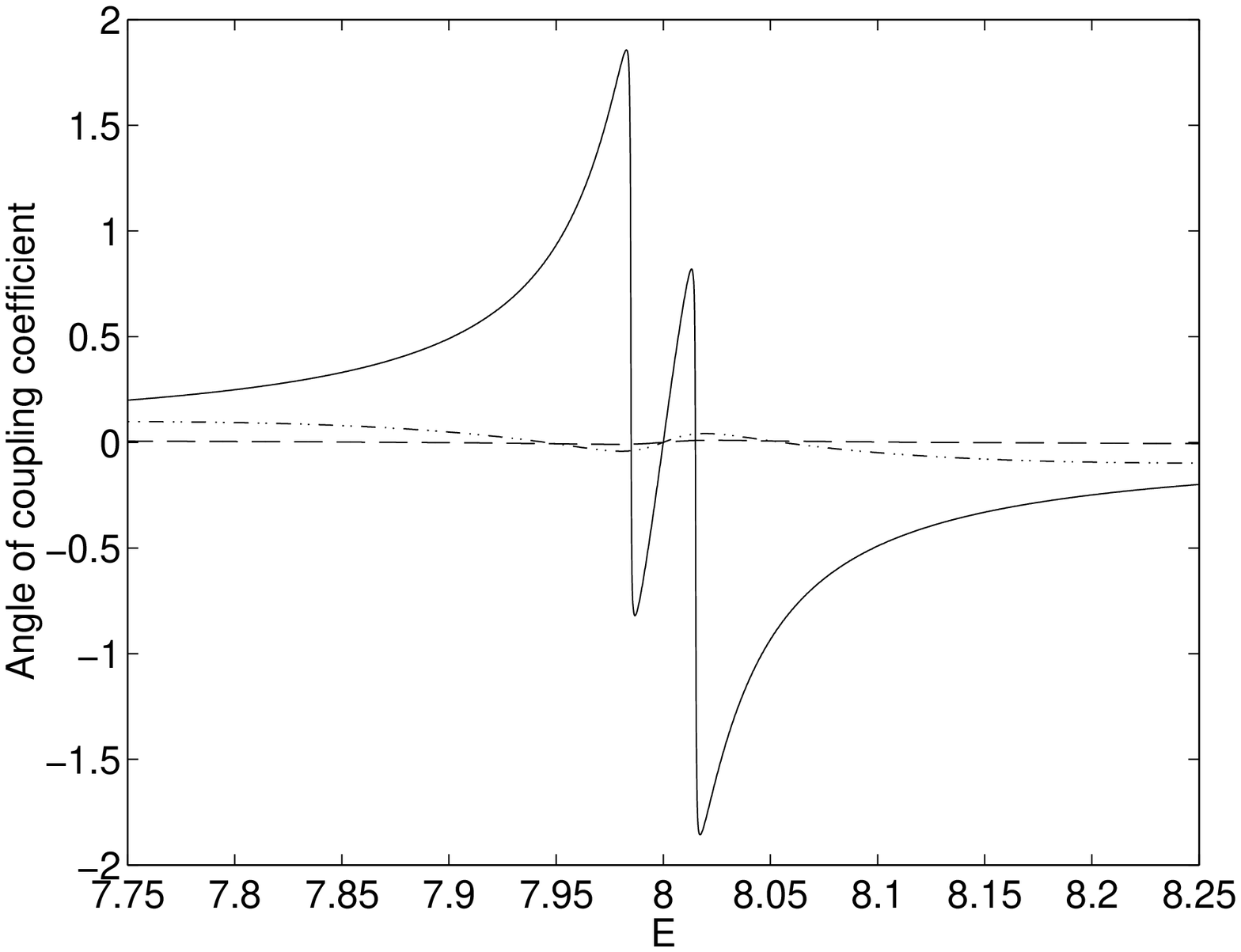}
\end{center}
\vspace*{.5cm}
\caption{
Cross section (top) with three resonance states at $\TE_1=7.99, ~\TE_2=8.01,
~\TE_3=8.0$ and $\TG_1 = \TG_2 = 0.05$, ~$\TG_3=0.05$ (full curve),
$\TG_3=1.0$ (dash-dotted curve), and $\TG_3=5.0$ (dashed curve).
Coupling coefficient (middle and bottom)
$\tilde W_3$ of the resonance state in the middle of the
spectrum.
$|\TW_3|$ is multiplied by 100 (full curve),
5 (dash-dotted curve) and 1 (dashed curve). 
Cross section and energy are given in arbitrary units.
}
\label{fig4}
\end{figure}

\begin{figure}
\begin{center}
\includegraphics[width=8cm,angle=0]{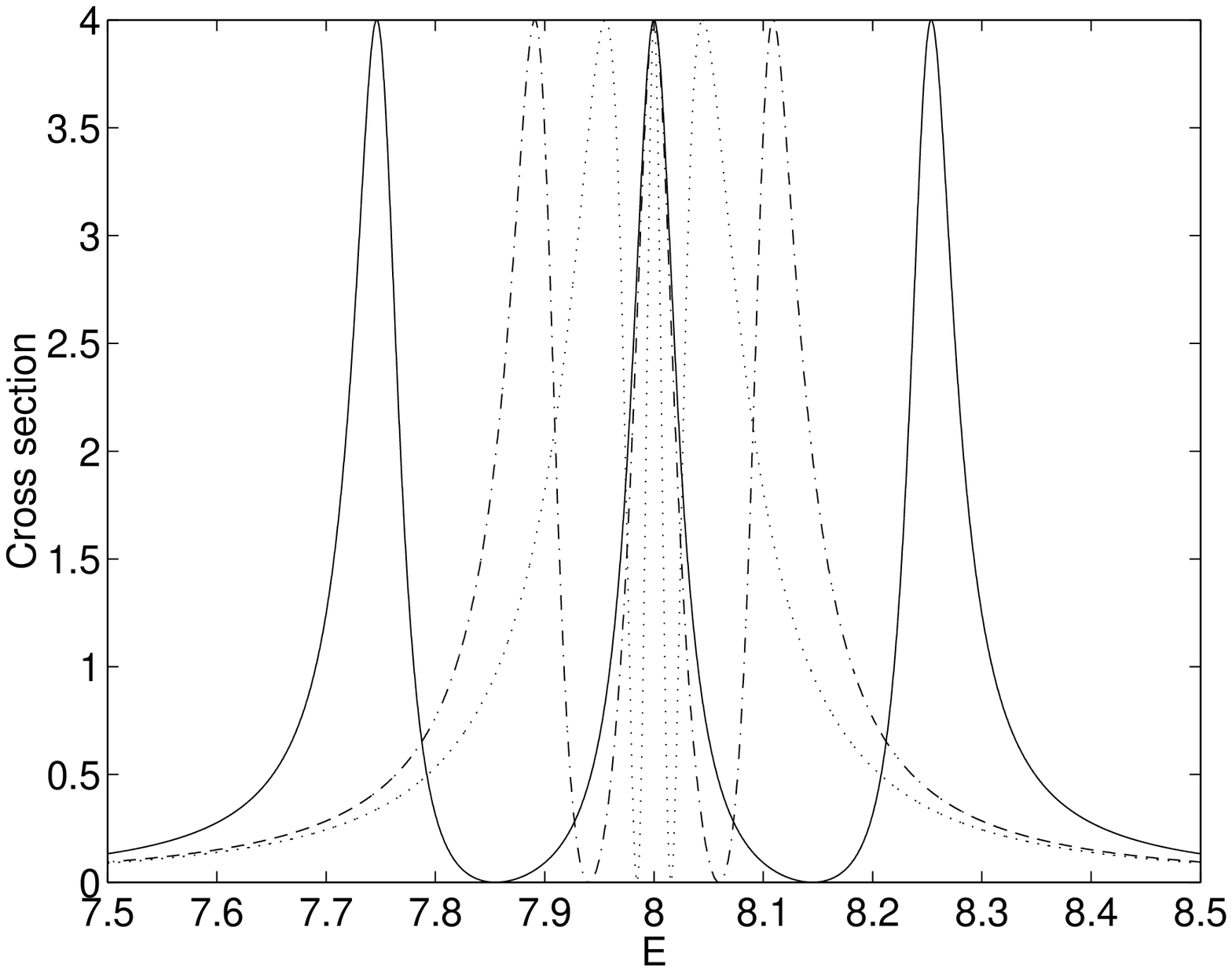} \\
\includegraphics[width=8cm,angle=0]{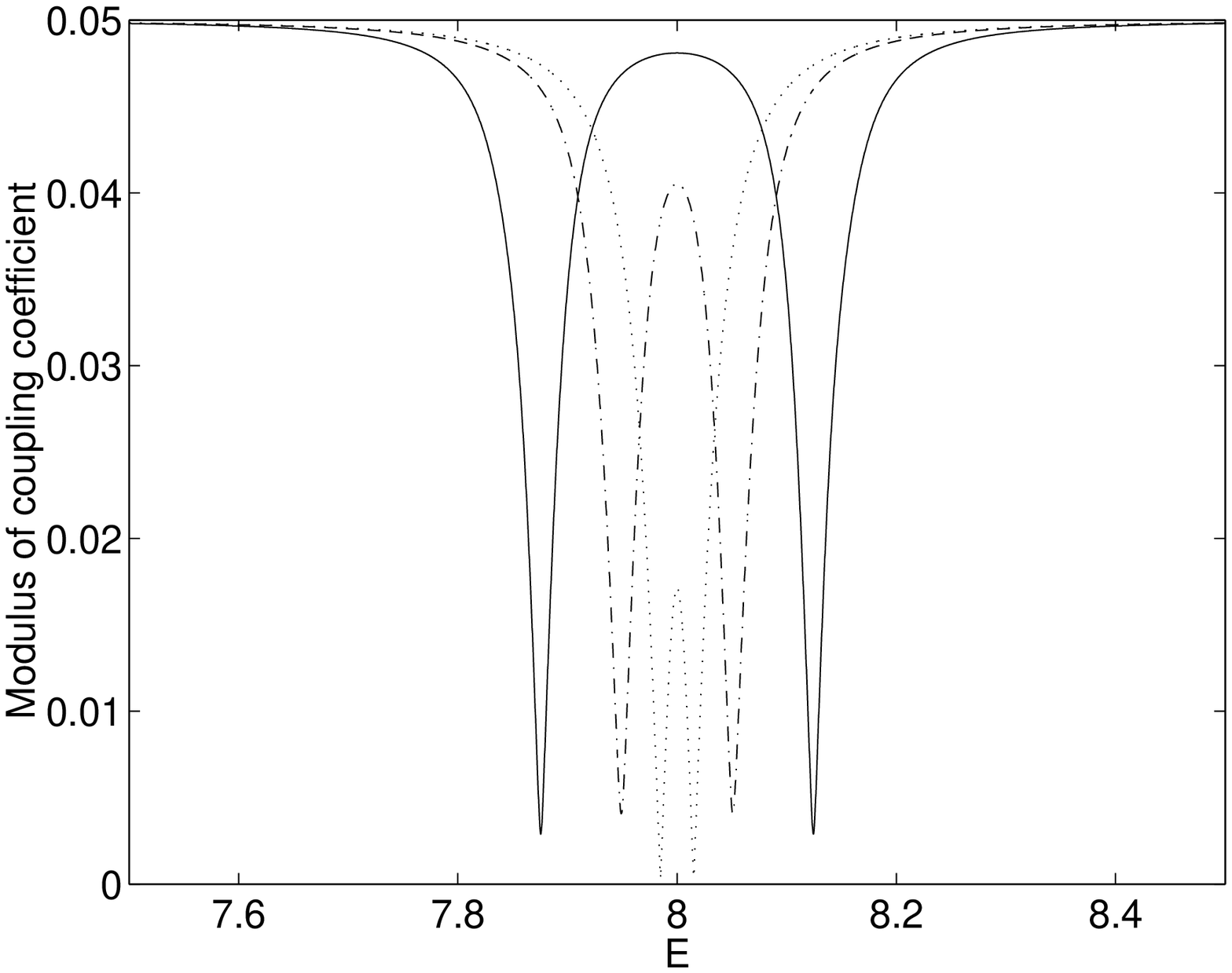} \\
\includegraphics[width=8cm,angle=0]{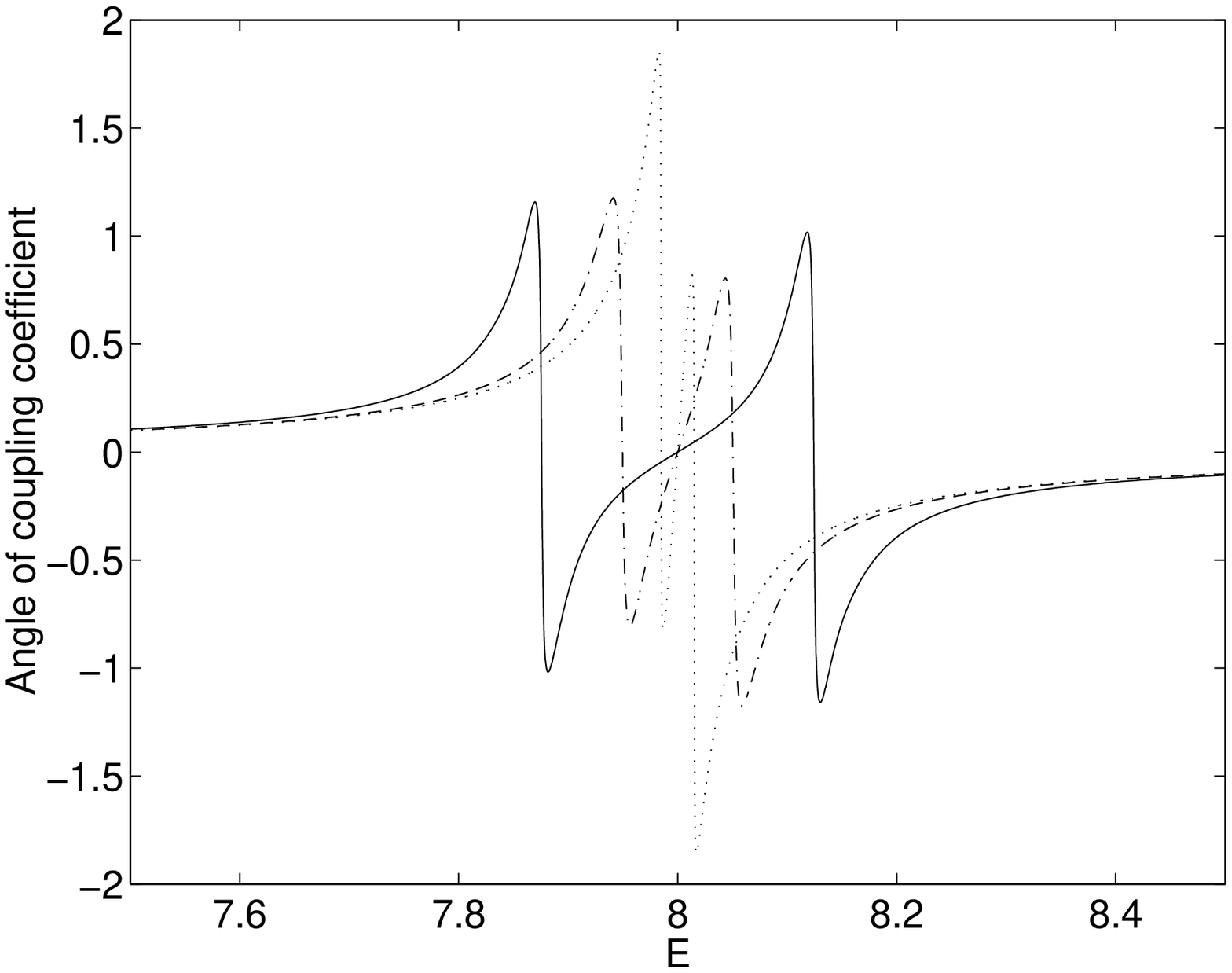} 
\end{center}
\vspace*{.5cm}
\caption{
Cross section (top) with three resonance states
of widths  $\TG_1 = \TG_2  =\TG_3=0.05$
at  $\TE_1=7.99, ~\TE_2=8.01, ~\TE_3=8.0$ (dotted curve), 
~$\TE_1=7.9, ~\TE_2=8.1, ~\TE_3=8.0$ (dash-dotted curve), 
~$\TE_1=7.75, ~\TE_2=8.25, ~\TE_3=8.0$ (full curve). 
Coupling coefficient $\tilde W_3$ (middle and bottom)
of the resonance state in the middle of the spectrum.
Cross section and energy are given in arbitrary units.
}
\label{fig5}
\end{figure}

\begin{figure}
\begin{center}
\includegraphics[width=8cm,angle=0]{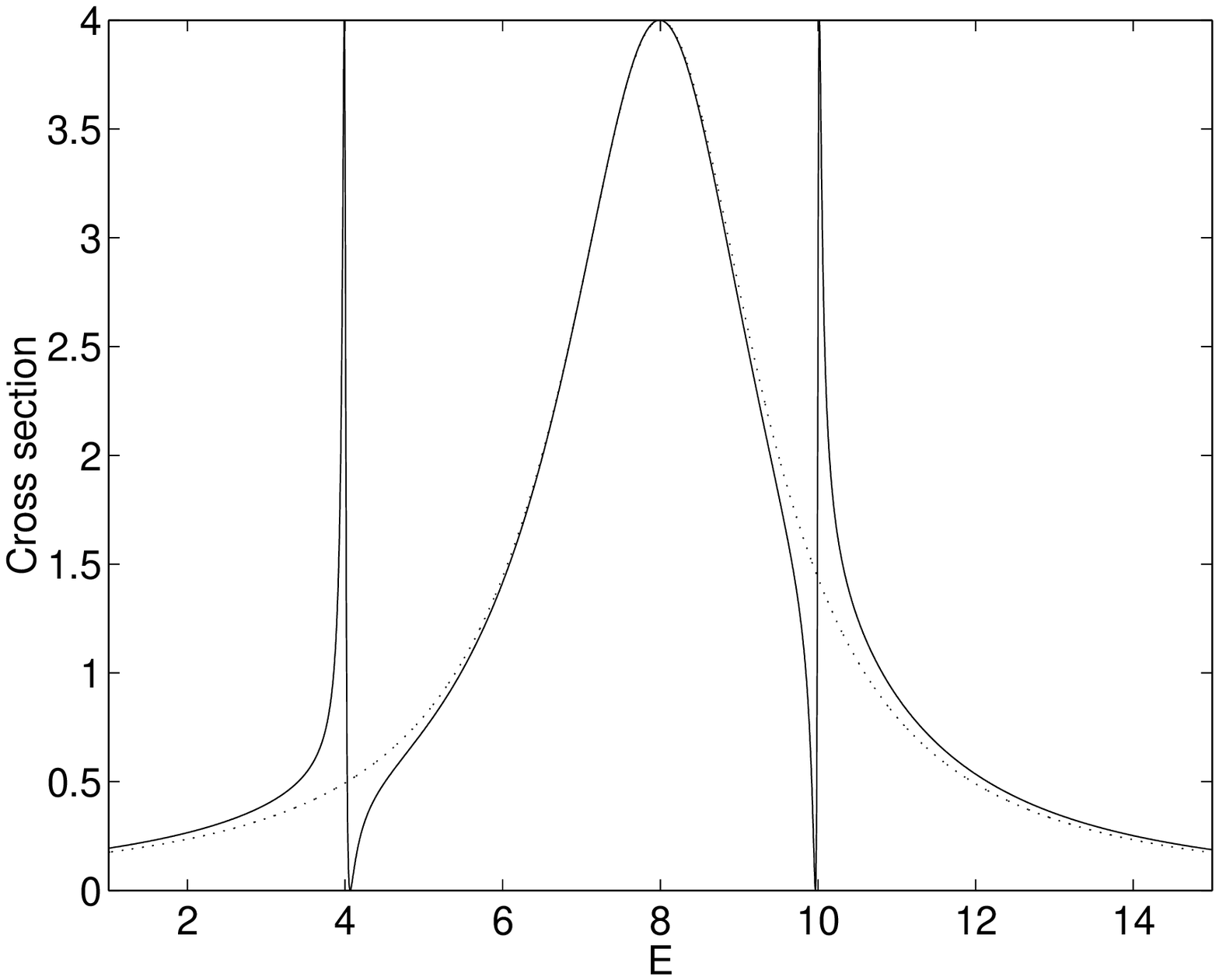}\\
\includegraphics[width=8cm,angle=0]{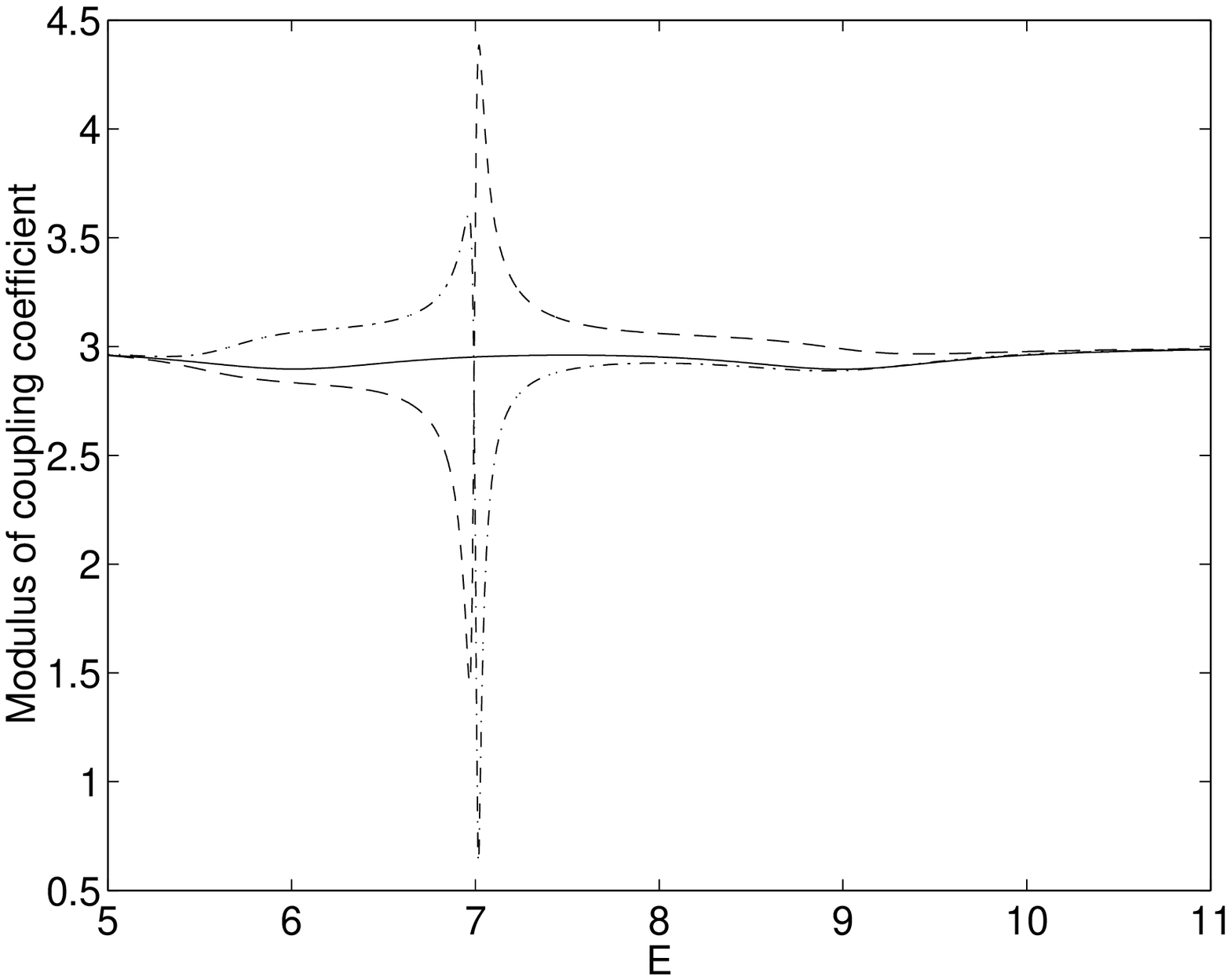}\\
\includegraphics[width=8cm,angle=0]{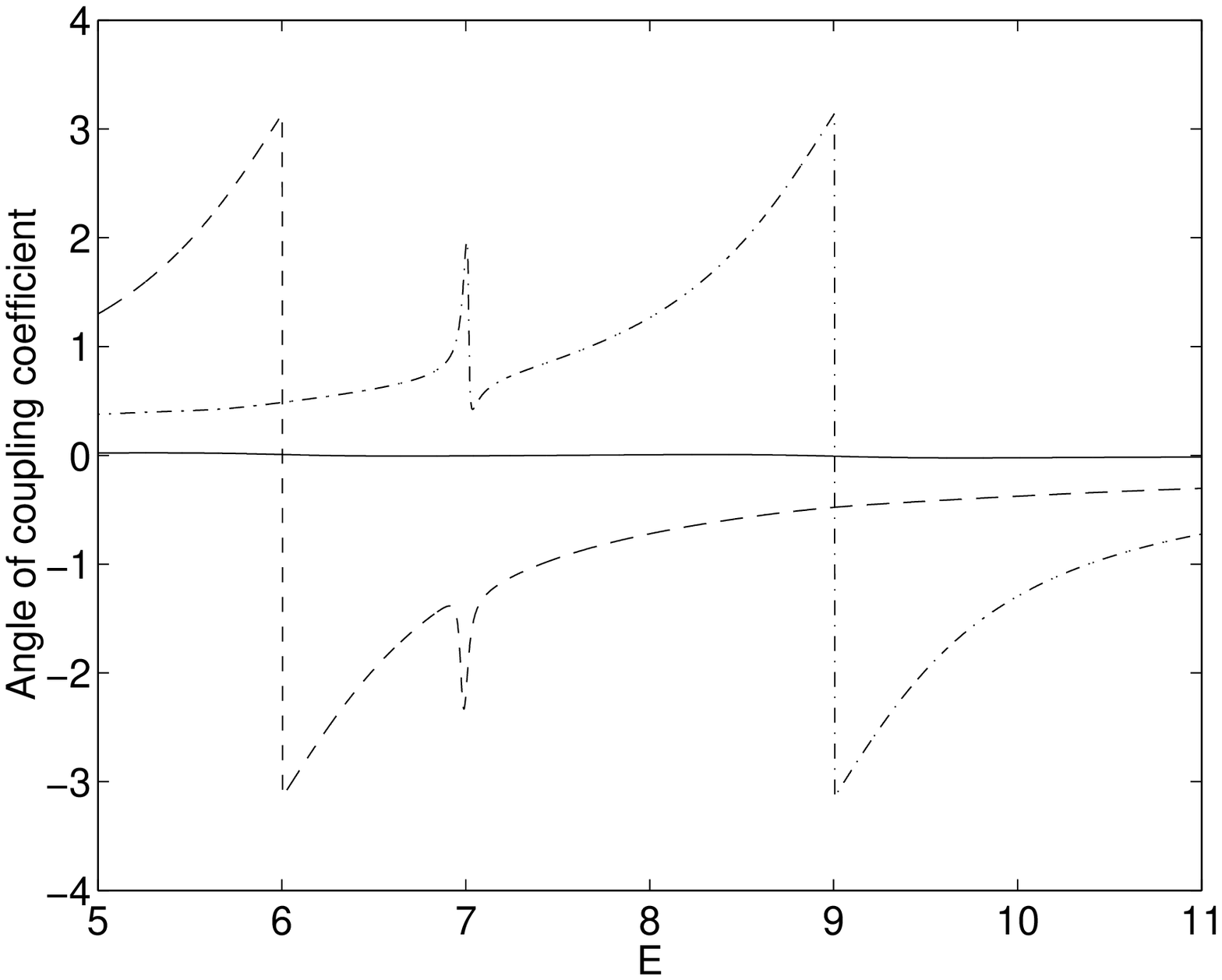}
\end{center}
\vspace*{.5cm}
\caption{
Cross section (top) with one  resonance state of width 
$\TG_3=3.0$ at $\TE_3=8.0$
(dotted curve) and with two additional  resonance states
of widths  $\TG_1 = \TG_2 = 0.05$
at  $\TE_1=4.0$ and $\TE_2=10.0$, respectively (full curve). 
Coupling coefficients (middle and bottom) 
$\tilde W_1$ (dashed), ~$\TW_2$ (dash-dotted) and 
$\TW_3$ (full curve) of the three resonance states.
$|\TW_1|$ and $|\TW_2|$ are multiplied by 60.
Cross section and energy are given in arbitrary units. Note the different 
energy scales for the cross section and the coupling coefficients.
}
\label{fig6}
\end{figure}

\begin{figure}
\begin{center}
\includegraphics[width=8cm,angle=0]{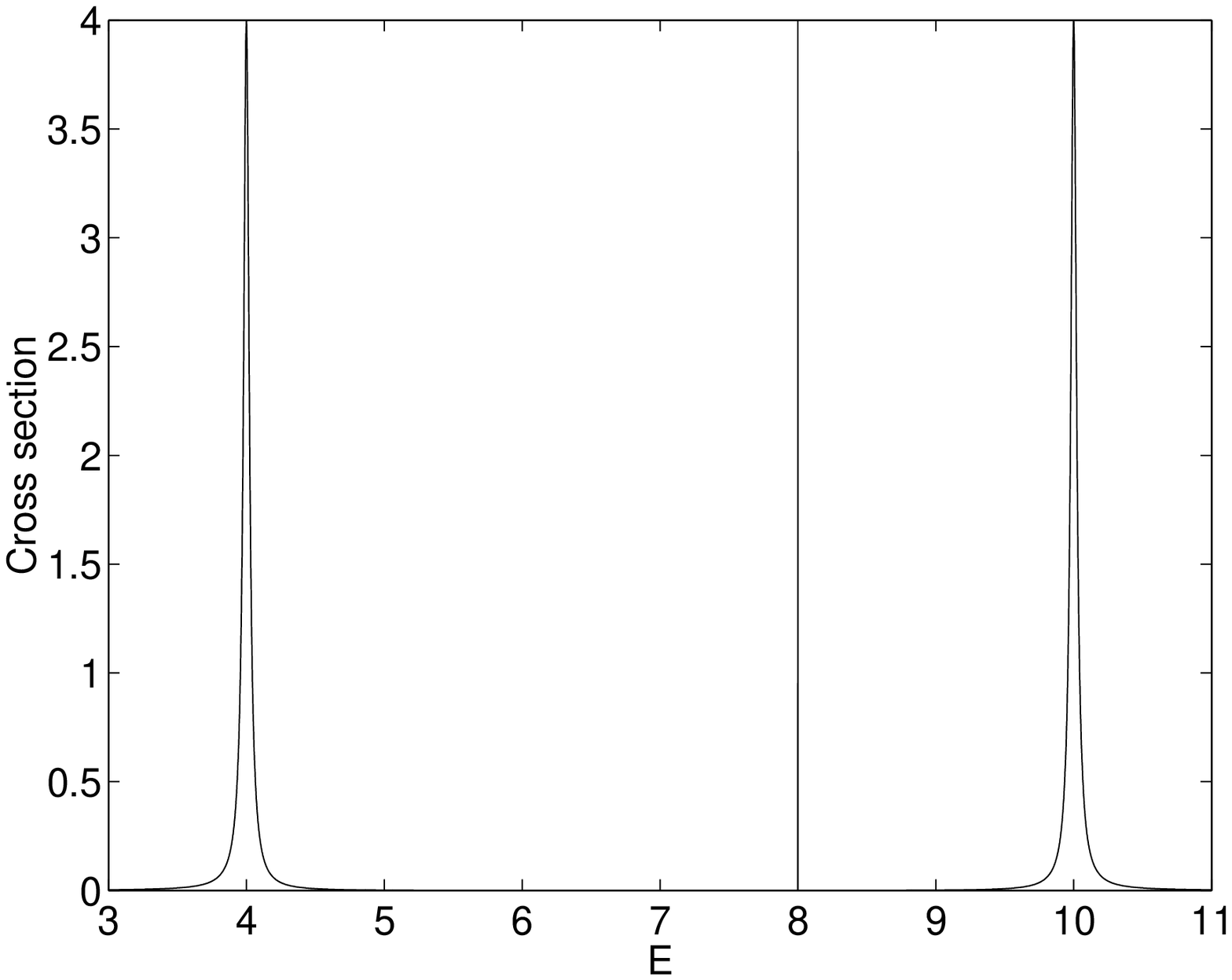}\\
\includegraphics[width=8cm,angle=0]{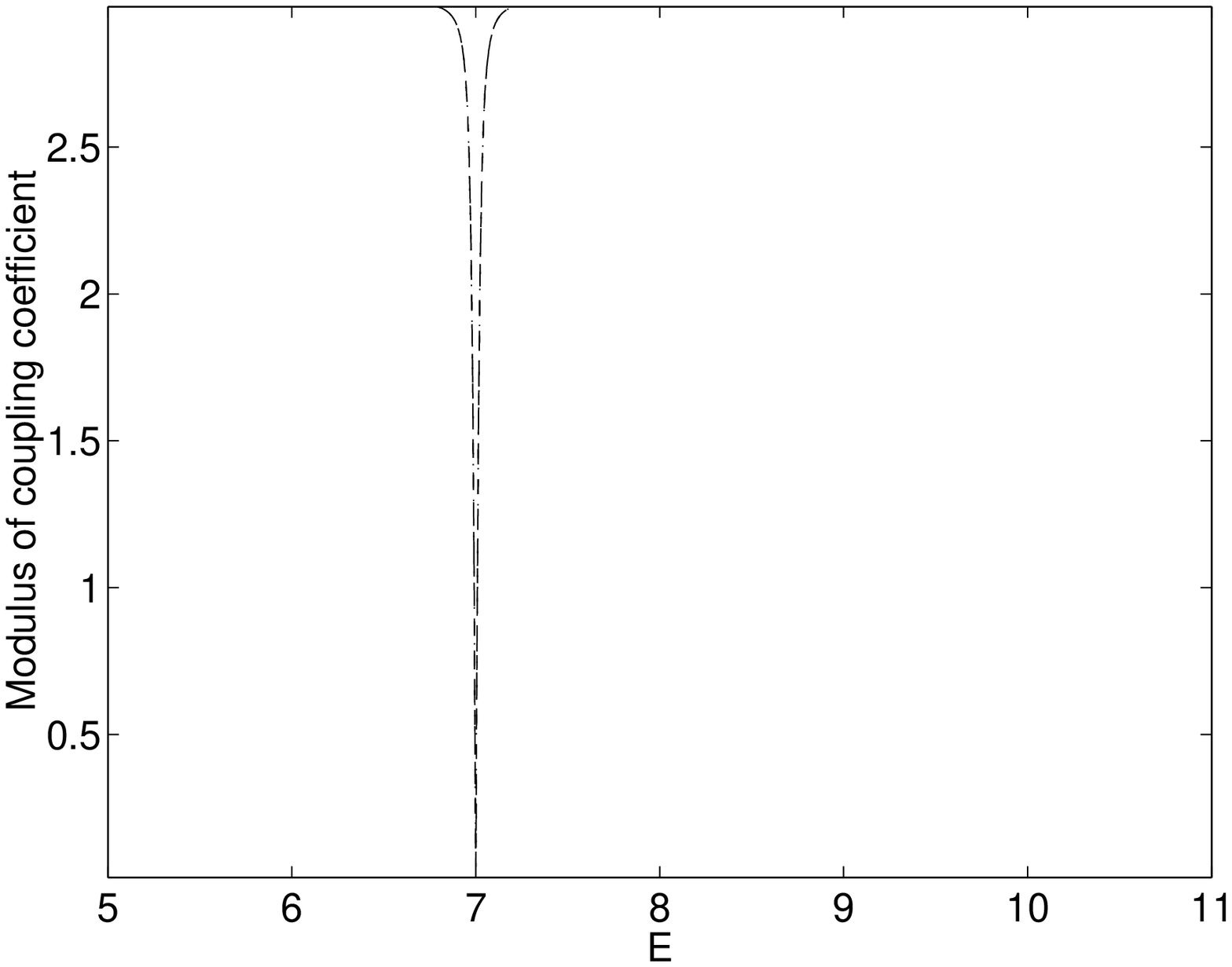} \\
\includegraphics[width=8cm,angle=0]{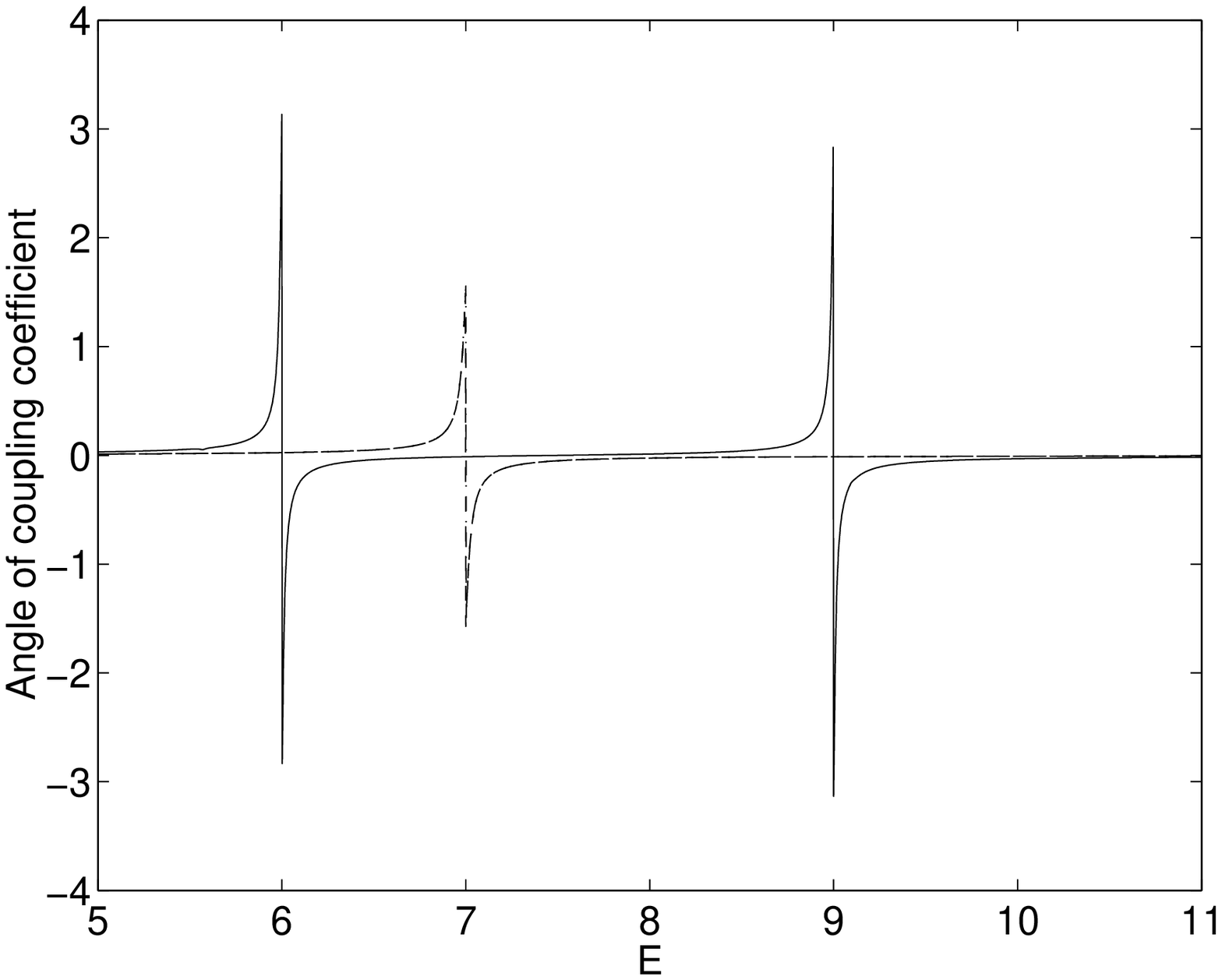}
\end{center}
\vspace*{.5cm}
\caption{
The same as  Fig. \ref{fig6}, but $\TG_3=3 \cdot 10^{-5}$.
}
\label{fig7}
\end{figure}


\begin{thebibliography}{99}
\bibitem{comment}
The interpretation of the experimental data given here on the basis of
Eqs. (\ref{wf2}) and (\ref{wf3}) differs from that given in \cite{demb2}. The
authors of \cite{demb2} consider the reduction of the dimension of the Hilbert 
space  at  $a=a^{\rm cr}$ (''defect of the Hilbert space'') that is caused 
by the coalescence of two eigenmodes. 
In the case studied,  the number of states is assumed to be reduced  from 2 
to 1 at $a^{\rm cr}$. This reduction of the number of states  
is suggested to be responsible for the
reduction of $\Delta$ from $\pi$ to $\pi/2$ at   $a^{\rm cr}$.
In difference to that interpretation, the reduction of $\Delta$ from $\pi$ 
to $\pi/2$ is caused, in the formalism given in
the present paper, first of all by the different degree
of mixing of the two states according to (\ref{wf3}). It takes place smoothly
in a certain region around $a^{\rm cr}$ in agreement with the experimental
data published in  \cite{demb2}.  In the present formalism, diabolic
points and exceptional points are related to one another. Nevertheless, they
are not the same in contrast to the statement given in reference [5] of 
\cite{demb2}.
\end{thebibliography}
\end{document}